\documentclass[review]{elsarticle}

\usepackage{algorithm}
\usepackage{algorithmic}
\usepackage{amsmath,amssymb,amsthm,lipsum}
\usepackage{graphicx}
\usepackage{textcomp}
\usepackage{stfloats}
\usepackage{booktabs}
\usepackage{bm}
\usepackage{subfigure}
\usepackage{appendix}

\theoremstyle{Theorem}
\newtheorem{theorem1}{Theorem}

\newtheorem{theorem6}[theorem1]{Theorem}
\newtheorem{theorem7}[theorem1]{Theorem}
\newtheorem{theorem8}[theorem1]{Theorem}

\theoremstyle{remark}
\newtheorem{rmk1}{Remark}
\newtheorem{rmk2}[rmk1]{Remark}
\newtheorem{rmk3}[rmk1]{Remark}
\newtheorem{rmk4}[rmk1]{Remark}
\newtheorem{rmk5}[rmk1]{Remark}
\newtheorem{rmk6}[rmk1]{Remark}

\theoremstyle{Definition}
\newtheorem{def1}{Definition}
\newtheorem{def2}[def1]{Definition}
\newtheorem{def3}[def1]{Definition}
\newtheorem{def4}[def1]{Definition}

\theoremstyle{Lemma}
\newtheorem{lemma1}{Lemma}

\newtheorem{lemma8}[lemma1]{Lemma}
\newtheorem{lemma9}[lemma1]{Lemma}

\newtheorem{lemma11}[lemma1]{Lemma}

\theoremstyle{Corollary}
\newtheorem{Corollary1}{Corollary}

\newtheorem{Corollary3}[Corollary1]{Corollary}
\newtheorem{Corollary4}[Corollary1]{Corollary}

\begin{document}

\begin{frontmatter}

\title{Compressive Spectrum Sensing Using \\
Blind-Block Orthogonal Least Squares}

\author[mymainaddress]{Liyang~Lu}

\author[mymainaddress]{Wenbo~Xu\corref{mycorrespondingauthor}}
\cortext[mycorrespondingauthor]{Corresponding authors:}
\ead{xuwb@bupt.edu.cn}

\author[mysecondaryaddress]{Yue~Wang\corref{mycorrespondingauthor}}
\ead{ywang56@gmu.edu}

\author[mysecondaryaddress]{Zhi~Tian}

\address[mymainaddress]{Key Lab of Universal
Wireless Communications, Ministry of Education, Beijing University of
Posts and Telecommunications}
\address[mysecondaryaddress]{Department of Electrical and Computer
Engineering, George Mason University, Fairfax, VA}

\begin{abstract}
Compressive sensing (CS) has recently emerged as an extremely efficient technology of the wideband spectrum sensing. In compressive spectrum sensing (CSS), it is necessary to know the sparsity or the noise information in advance for reliable reconstruction. However, such information is usually absent in practical applications. In this paper, we propose a blind-block orthogonal least squares-based compressive spectrum sensing (B-BOLS-CSS) algorithm, which utilizes a novel blind stopping rule to cut the cords to these prior information. Specifically, we first present both the noiseless and noisy recovery guarantees for the BOLS algorithm based on the mutual incoherence property (MIP). Motivated by them, we then formulate the blind stopping rule, which exploits an $\ell_{2,\infty}$ sufficient statistic to blindly test the support atoms in the remaining measurement matrix. We further evaluate the theoretical performance analysis of the holistic B-BOLS-CSS algorithm by developing a lower bound of the signal-to-noise ratio (SNR) to ensure that the probability of exact recovery is no lower than a given threshold. Simulations not only demonstrate the improvement of our derived theoretical results, but also illustrate that B-BOLS-CSS works well in both low and high SNR environments.
\end{abstract}


\end{frontmatter}


\section{Introduction}

Spectrum sensing is a key issue for efficient utilization of spectrum resources in the cognitive radios (CR) \cite{24,25,21,18,22,23}. It dynamically provides the spectrum occupancy status for the secondary users (SUs), so as to avoid harmful interference to primary users (PUs) due to SUs' access. Since wider spectrum range provides more access opportunities for SUs, wideband spectrum sensing has received much recent attention. One challenge in the implementation of wideband spectrum sensing lies in its prohibitive sampling rate. As a real-time processing desired problem, wideband spectrum sensing thus calls for recovery technology with fast implementation. A remarkable choice is arguably the compressive sensing (CS), which is used instead of traditional Nyquist sampling to reduce hardware complexity and cost.

Although the compressive spectrum sensing (CSS) is able to successfully sense the wideband spectrum with the sampling rate close to the information rate, it requires sparsity or noise information in advance. Otherwise, the CS algorithms cannot stop iteration in time and cause performance loss. To address this issue, a well-known method is the so-called blind compressive sensing (BCS) \cite{36}. The primitive BCS contains the sparse coding and the measurement matrix update steps, which is similar to K-SVD dictionary learning \cite{52}. Its later extension \cite{37} considers the issue of simultaneous signal reconstruction with the assumption that the original signals come from the union of a few disjoint subspaces. Other application variants can be referred to \cite{44,45,43}. The drawback of blind compressed sensing lies in the complex dictionary learning procedure and the subspace constraint. This drawback limits the applicability of blind compressive sensing in wideband spectrum sensing.

Recently, the authors in \cite{4} develop a blind stopping rule for orthogonal matching pursuit (OMP) by blindly testing the support signal in the remaining measurement matrix per iteration, lessening the dependence on prior information. The recovery performance of OMP with this blind stopping rule is comparable to the alternative with prior information. Hence, it is indeed a favorable choice for wideband spectrum sensing. Other researches, such as \cite{21,54}, address the unknown sparsity issue effectively by utilizing statistical analysis to estimate the sparsity level, and hence are applied to spectrum sensing.

Though OMP has been well developed for spectrum sensing, the utilization of other greedy algorithms remain barely explored
in this field. Orthogonal least squares (OLS) \cite{10,8,2} is a typical CS greedy algorithm with better convergence property than OMP \cite{9,15}. Meanwhile, OLS depends less on the amplitude distribution of nonzero entries. These advantages motivate us to consider OLS as a more reliable candidate for spectrum sensing.

Further, it has been pointed that PUs signals exhibit block sparsity, i.e., the nonzero entries appear in blocks \cite{3,55}.
Researches have proved that exploiting block sparsity during sparse signal recovery usually brings great performance gain and better convergence \cite{28,31,32,39,42,3,29,30,35,40}. Such gain has also been verified in a number of applications including block spectrum sensing \cite{18,27}, face recognition \cite{34,19,41} and so on.
In \cite{56,57}, the authors propose blind-block sparse recovery methods based on Bayesian framework of neighborhood statistics, which do not need any prior information of the sparsity structure. However, they are constrained by the huge consumption of computing resources. In \cite{58}, a deterministic iterative neighborhood based blind-block algorithm, which is better than those in \cite{56,57} in both running time and recovery performance, is provided via a predefined sparsity inducing function. This function excessively relies on the selection of the neighborhood radius and thus leads to the instability of the algorithm.
Therefore, to improve the efficiency and stability of the CSS procedure while considering block sparsity, it is necessary to develop a practical blind-block sensing algorithm.

In this paper we propose a novel blind-block orthogonal least squares-based compressive spectrum sensing algorithm (B-BOLS-CSS) for realizing the real-time and accurate CSS. Specifically, we first formulate the extended exact recovery conditions (ERCs) for BOLS by tightening the matrix eigenvalue bounds via mutual incoherence property (MIP) \cite{3,2}, which is a widely used metric for sparse signal recovery, in the noiseless scenario. Then, based on the derived recovery conditions, the blind stopping rule for BOLS algorithm is developed, followed by its theoretical performance guarantees in the noisy scenario. The B-BOLS-CSS algorithm is proposed by combining the blind stopping rule and BOLS algorithm. The main contributions are given as follows.
\begin{enumerate}

    \item Both the noiseless and noisy recovery conditions for BOLS algorithm are derived, which are better than the existing ones. Specifically, we develop a tighter eigenvalue bounds than those in \cite{1} by utilizing the block orthogonality. Then, based on these eigenvalue bounds, an extended upper bound of the reconstructible sparsity level is derived, which is better than that in \cite{2} and acts as a solid foundation for the subsequent derivation of the blind stopping rule.

    \item We propose a blind stopping rule, which is incorporated with $\ell_{2,\infty}$ norm, for BOLS by using the aforementioned theoretical results. A novel B-BOLS-CSS algorithm is then proposed by combining this blind stopping rule and BOLS algorithm. The lower bound of the ${\rm SNR}_{\min}$ required for reliable recovery of B-BOLS-CSS is developed, which is lower than that in \cite{4}. These analyses reveal that B-BOLS-CSS algorithm performs better than the CSS using blind OMP algorithm.

    \item The simulations demonstrate the superiority of our derived theoretical results compared with the existing ones. Meanwhile, we evaluate the effectiveness and feasibility of our proposed B-BOLS-CSS algorithm. The performance of B-BOLS-CSS with unknown number of active PUs and the noise variance is close to that of CSS using the algorithms, e.g., conventional block OMP (BOMP) and BOLS, with these prior information, and is more robust than the one without utilizing block structure.

\end{enumerate}

The rest of this paper is organized as follows. In Section~ ${\rm\uppercase\expandafter{\romannumeral2}}$, we introduce notations, CSS model and basic definitions, which facilitate the subsequent study of theoretical analysis and algorithm proposal of B-BOLS-CSS in Section ${\rm \uppercase\expandafter{\romannumeral3}}$. We present simulation results in Section ${\rm \uppercase\expandafter{\romannumeral4}}$, followed by conclusions in Section ${\rm \uppercase\expandafter{\romannumeral5}}$.

\section{Preliminaries}
\subsection{Notations and Assumptions}
In this paper, we denote vectors by boldface lower-case letters, e.g., $\mathbf{r}$, and matrices by boldface upper-case letters, e.g., $\mathbf{D}$. The entry of vector $\mathbf{r}$ and matrix $\mathbf{D}$ are denoted as $r_i$ and $\mathbf{D}_{ij}$ respectively. $\mathbf{D}_i$ is the $i$-th column of $\mathbf{D}$. $\mathbf{D}^T$ represents the transpose of matrix $\mathbf{D}$. $\mathbf{D}\setminus \mathbf{D}_0$ represents the matrix whose atoms do not belong to $\mathbf{D}_0$ but belong to $\mathbf{D}$. If the quantity in $|\cdot|$ is a numerical value, $|\cdot|$ means its absolute value.
$\mathbf{D}_{\mathbf{S}^t}$ is a submatrix of $\mathbf{D}$ that contains the column set $\mathbf{S}^t$ selected during the $t$-th iteration. The symbol span$(\mathbf{D})$ represents the span of columns in $\mathbf{D}$.
During the iteration, if the selected matrix $\mathbf{D}_{\mathbf{S}^t}$ has full column rank, then $\mathbf{P}_{\mathbf{S}^t}=\mathbf{D}_{\mathbf{s}^t}\mathbf{D}_{\mathbf{s}^t}^{\dag}$ stands for the projection onto span$(\mathbf{D}_{\mathbf{s}^t})$, where $\mathbf{D}_{\mathbf{S}^t}^\dag=(\mathbf{D}_{\mathbf{s}^t}^T\mathbf{D}_{\mathbf{s}^t})^{-1}\mathbf{D}_{\mathbf{s}^t}^T$ is the pseudoinverse of $\mathbf{D}_{\mathbf{s}^t}$. $\mathbf{P}_{\mathbf{S}^t}^\bot=\mathbf{I}-\mathbf{P}_{\mathbf{S}^t}$ is the projection onto the orthogonal complement of span$(\mathbf{D}_{\mathbf{S}^t})$. The spectral norm of a matrix $\mathbf{D}$ is denoted by $\rho(\mathbf{D})=\sqrt{\lambda_{\max}(\mathbf{D}^T\mathbf{D})}$, where $\lambda_{\max}(\mathbf{A})$ is the largest eigenvalue of $\mathbf{A}$. In addition, throughout the paper, we assume the measurement matrix is normalized, i.e., the $\ell_2$-norm of each column in $\mathbf{D}$ is equal to 1.

\subsection{Compressive Spectrum Sensing}
The system model of CSS is given as follows:
\begin{equation}\label{CSmodel}
\mathbf{y}=\mathbf{D}\mathbf{x}+\mathbf{\epsilon},
\end{equation}
where $\mathbf{y}\in \mathcal{R}^{m\times1}$ is the low-dimensional measurement vector, $\mathbf{D}\in \mathcal{R}^{m\times n}$ is the measurement matrix, $\mathbf{x}\in \mathcal{R}^{n\times1}$ is the spectrum vector with $n>m$ and $\mathbf{\epsilon}\in \mathcal{R}^{m\times1}$ represents the measurement noise. The number of the nonzero elements $K$ in the spectrum vector $\mathbf{x}$ is called sparsity. The spectrum sensing algorithms are designed to accurately recover $\mathbf{x}$ from the given measurement vector $\mathbf{y}$. Then, the SUs can perform interference free access with the help of the recovered $\hat{\mathbf{x}}$.

It has been pointed that the spectrum $\mathbf{x}$ is usually block sparse. Then, block compressed sensing (BCS) is proposed in \cite{3} to recover the signal. The number of nonzero blocks in sparse signal $\mathbf{x}$ is represented by $k$, i.e., block sparsity, in this paper. Let $d$ denote the block length, $N_B$ denote the total number of blocks in $\mathbf{x}$, and the block sparse spectrum $\mathbf{x}$ is defined as
\begin{equation}\label{sparsex}
\mathbf{x}=[\underbrace{x_1\cdots  x_d }_{\mathbf{x}^T[1]} \underbrace{x_{d+1}\cdots x_{2d}}_{\mathbf{x}^T[2]}\cdots \underbrace{x_{n-d+1}\cdots x_{n}}_{\mathbf{x}^T[N_B]}]^T,
\end{equation}
where $n=N_Bd$ and $\mathbf{x}[i]\in \mathcal{R}^{d\times1}$ denotes the $i$-th block of $\mathbf{x}$. The measurement matrix can be rewritten as a concatenation of $N_B$ column blocks, i.e.,
\begin{equation}\label{matrixblock}
\mathbf{D}=[\underbrace{\mathbf{D}_1\cdots  \mathbf{D}_d }_{\mathbf{D}[1]} \underbrace{\mathbf{D}_{d+1}\cdots \mathbf{D}_{2d}}_{\mathbf{D}[2]}\cdots \underbrace{\mathbf{D}_{n-d+1}\cdots \mathbf{D}_{n}}_{\mathbf{D}[N_B]}],
\end{equation}
where $\mathbf{D}[i]\in \mathcal{R}^{m\times d}$ is the $i$-th block of $\mathbf{D}$.

\subsection{Definitions}
In this subsection, we give the definitions of matrix coherence, block-coherence and sub-coherence.

\begin{def1}
The matrix coherence of a matrix $\mathbf{D}$, which measures the similarity of its entries, is defined as:
\begin{equation}\label{definitionmatrixcoherence}
\mu=\max_{i,j\neq i}|\mathbf{D}_i^T\mathbf{D}_j|.
\end{equation}
\label{def1}
\end{def1}

\begin{def2}
The block-coherence of $\mathbf{D}$ is defined as:
\begin{equation}\label{definitionmatrixcoherence2}
\mu_B=\max_{i,j\neq i}\frac{||\mathbf{M}[i,j]||_2}{d},
\end{equation}
where $\mathbf{M}[i,j]=\mathbf{D}^T[i]\mathbf{D}[j]$.
\label{def2}
\end{def2}

\begin{def3}
The sub-coherence of $\mathbf{D}$ is given by:
\begin{equation}\label{definitionmatrixcoherence3}
\nu=\max_l\max_{i,j\neq i}|\mathbf{D}_i^T\mathbf{D}_j|,\,\mathbf{D}_i,\mathbf{D}_j\in \mathbf{D}[l].
\end{equation}
\label{def3}
\end{def3}

\begin{def4}
The ${\rm SNR}$ is defined as
\begin{equation}\label{conventionalsnr}
{\rm SNR}=\frac{\mathbb{E}(||\mathbf{D}\mathbf{x}||_2^2)}{\mathbb{E}(||\mathbf{\epsilon}||_2^2)},
\end{equation}
where $\mathbb{E}(\cdot)$ represents the expectation of its objective.
The component ${\rm SNR}$ is given by
\begin{equation}\label{componentsnr}
{\rm SNR}_q=\frac{||\mathbf{x}_q\mathbf{D}_q||_2^2}{M\sigma^2} ,\;q=1,2,\cdots,n,
\end{equation}
and
the minimum component ${\rm SNR}_{\min}$ is the minimum value of the component ${\rm SNR}$s \cite{4}.
\label{def4}
\end{def4}

\section{Recovery Analysis and Blind Stopping Rule for BOLS}
In this section, we first derive the extended MIP-based condition for ERC of BOLS algorithm. Then, based on the derived condition, we present the blind stopping rule for BOLS and provide the B-BOLS-CSS algorithm.

\subsection{MIP Condition in the Noiseless Scenario}

Without loss of generality, assume that the first $kd$ entries of the sparse spectrum $\mathbf{x}$ are nonzero and the set containing selected indices is $\mathbf{S}^t=\{1,\cdots,td\}$. Then, in the $t+1$-th iteration, define
$\mathbf{D}_0=[\mathbf{D}_1, \mathbf{D}_2,\cdots,\mathbf{D}_{kd}]=[\mathbf{D}[1], \mathbf{D}[2],\cdots,\mathbf{D}[k]]$,
$\mathbf{D}_{0\backslash\mathbf{S}}=\mathbf{D}_0\backslash\mathbf{D}_{\mathbf{S}^t}$
and
$\overline{\mathbf{D}}_0=\mathbf{D}\backslash\mathbf{D}_0$. $\mathbf{R}_{0\backslash\mathbf{S}}$ and $\overline{\mathbf{R}}_0$ corresponding to $\mathbf{D}_{0\backslash\mathbf{S}}$ and $\overline{\mathbf{D}}_0$ are defined by
\begin{equation}\label{Rodefinition}
\begin{aligned}
\mathbf{R}_{0\backslash\mathbf{S}}= f(\mathbf{D}_{0\backslash\mathbf{S}})
=\left[
\begin{matrix}

   \frac{1}{||\mathbf{P}^\bot_{\mathbf{S}^{t}}\mathbf{D}_1||_2} &  &\mathbf{0}\\

     &       \ddots&\\
     \mathbf{0}&&\frac{1}{||\mathbf{P}^\bot_{\mathbf{S}^{t}}\mathbf{D}_{(k-t)d}||_2}
\end{matrix}
\right]
\end{aligned}
\end{equation}
and $\overline{\mathbf{R}}_0= f(\overline{\mathbf{D}}_0)$.

The condition
\begin{equation}\label{ercdefined}
\gamma=\rho_c[(\mathbf{D}_{0\backslash\mathbf{S}}\mathbf{Q}_{0\backslash\mathbf{S}})^{\dag}(\overline{\mathbf{D}}_{0\backslash\mathbf{S}}\overline{\mathbf{Q}}_{0\backslash\mathbf{S}})]<1
\end{equation}
is called the ERC for BOLS algorithm \cite{2}, where
$\rho_c(\mathbf{A})=\max\limits_{j}\sum\limits_{i}\rho(\mathbf{A}[i,j])$
and $\mathbf{A}[i,j]$ is the $(i,j)$-th block of $\mathbf{A}$.
 In the following, we present the detailed analysis for the improved MIP-based sufficient condition for the establishment of this ERC.

\begin{lemma8}
For $\mu_B$ and $\nu$ in Definitions \ref{def2} and \ref{def3}, suppose $(k-1)d\mu_B<1$ and $\nu=0$, then $1-(k-1)d\mu_B\leq\lambda_{\min}\leq\lambda_{\max}\leq1+(k-1)d\mu_B$, where $d$ is the block length, $\lambda_{\min}$ and $\lambda_{\max}$ denote the minimum and maximum eigenvalues of $\mathbf{D}_0^T\mathbf{D}_0$.
\label{lemma8}
\end{lemma8}

\begin{proof}
See Appendix \ref{proofoflemma8}.
\end{proof}

\begin{rmk3}
Our derived result in Lemma \ref{lemma8} is much tighter due to the utilizations of block sparsity and block orthogonality than the well-known result, i.e., Lemma 2 in \cite{1}, $1-(kd-1)\mu\leq\lambda_{\min}\leq\lambda_{\max}\leq1+(kd-1)\mu$. This is because the block-coherence satisfies $\mu_B\leq\mu$ and $d\mu_B\geq\mu$ \cite{2}, which indicates that $1-(k-1)d\mu_B\geq1-(kd-1)\mu$ and $1+(k-1)d\mu_B\leq1+(kd-1)\mu$.
\label{rmk3}
\end{rmk3}

\begin{rmk2}
Denote the compression rate as a fixed constant $\tau$.
Then $\lim\limits_{m/n=\tau;m,n\rightarrow\infty}\nu=0$, which indicates the assumption $\nu=0$ in Lemma~\ref{lemma8} is reasonable in high-dimensional application, e.g., wideband spectrum sensing.
\end{rmk2}

To compare the derived ranges of eigenvalues using block-structure property with the one using conventional structure, we give the following Corollary \ref{corollary3}.

\begin{Corollary3}
For $\mu_B$ and $\nu$ in Definitions \ref{def2} and \ref{def3}, suppose $(K-d)\mu_B<1$ and $\nu=0$, then $1-(K-d)\mu_B \leq\lambda_{\min}\leq\lambda_{\max}\leq1+(K-d)\mu_B$, where $d$ is the block length.
\label{corollary3}
\end{Corollary3}

\begin{rmk4}
Corollary \ref{corollary3} is derived by replacing $kd$ with $K$ in Lemma~\ref{lemma8}. Compared with Lemma 2 in \cite{1}, i.e.,
\begin{equation}\label{lemma2in35}
1-(K-1)\mu \leq\lambda_{\min}\leq\lambda_{\max}\leq1+(K-1)\mu,
\end{equation}
the bounds of eigenvalues in Corollary \ref{corollary3} are tighter since $d\geq1$ and $\mu_B\leq\mu$. This tightness can be attributed to the consideration of block structure \cite{3,2}.
\label{rmk4}
\end{rmk4}

In order to present the extended MIP-based condition for ERC of the BOLS algorithm, we first provide the following lemma, which describes a new lower bound for $||\mathbf{P}_t^{\bot}\mathbf{D}_i||_2$.

\begin{lemma1}
For $\mu$, $\mu_B$ and $\nu$ in Definitions \ref{def1}, \ref{def2} and \ref{def3}, suppose $(k-1)d\mu_{B}<1$ and $\nu=0$, then $||\mathbf{P}_{\mathbf{S}^t}^{\bot}\mathbf{D}_i||_2\geq\sqrt{1/\mathcal{B}}$ for $i\in\{1,2,\cdots,n\}\backslash \mathbf{S}^t$,
where $\mathcal{B}=\Big(1-\frac{kd\mu^2(1+(k-1)d\mu_B)}{(1-(k-1)d\mu)^2}\Big)^{-1}$.
\label{lemma1}
\end{lemma1}

\begin{rmk1}
There are two existing lower bounds for $||\mathbf{P}_{\mathbf{S}^t}^{\bot}\mathbf{D}_i||_2$. Those are,
the bound using Lemmas 2 and 5 in \cite{1}:
\begin{equation}\label{existingbound1}
||\mathbf{P}_{\mathbf{S}^t}^{\bot}\mathbf{D}_i||_2\geq\sqrt{1-kd\mu},
\end{equation}
and the bound in our previous work \cite{2}:
\begin{equation}\label{existingbound2}
||\mathbf{P}_{\mathbf{S}^t}^{\bot}\mathbf{D}_i||_2\geq\sqrt{1-\Big(\frac{\sqrt{1+(kd-1)\mu}\sqrt{kd\mu^2}}{1-(k-1)d\mu}\Big)^2}.
\end{equation}
Similar to the analysis of remark \ref{rmk3}, our derived result is tighter than these existing ones due to the extra analysis about block sparsity.
\end{rmk1}

The proof of Lemma \ref{lemma1} is similar to that of Lemma 3 in \cite{2} but with our derived Lemma \ref{lemma8} in this paper.
Based on Lemma \ref{lemma1}, the MIP-based sufficient condition for the ERC of the BOLS is presented in the following Theorem~\ref{theorem7}.

\begin{theorem7}
The ERC in (\ref{ercdefined}) is satisfied if
\begin{equation}\label{MIPERC}
k<(\sqrt[3]{-\mathcal{Q}/2+\sqrt{\Delta}}+\sqrt[3]{-\mathcal{Q}/2-\sqrt{\Delta}}-\beta/3\alpha),
\end{equation}
where
$\mathcal{Q}=\frac{27\alpha^2\delta-9\alpha\beta\omega+2\beta^3}{27\alpha^3}$, $\mathcal{P}=\frac{3\alpha\omega-\beta^2}{3\alpha^2}$, $\alpha=-d^3\mu_B^2\mu^2+3d^3\mu_B\mu^2$, $\beta=d^3\mu_B^2\mu^2+d^2\mu_B\mu^2-7d^3\mu_B\mu^2-6d^2\mu_B\mu-2d^2\mu^2$, $\omega=5d^3\mu_B\mu^2-2d^2\mu_B\mu^2+8d^2\mu_B\mu+4d^2\mu^2+2d\mu^2+3d\mu_B+4d\mu$, $\delta=-d^3\mu_B\mu^2-2d^2\mu_B\mu-2d^2\mu^2-d\mu_B-4d\mu-2$ and $\Delta=(\mathcal{Q}/2)^2+(\mathcal{P}/3)^3$.
\label{theorem7}
\end{theorem7}

\begin{proof}
By using Eqn. (100) in \cite{2} and Lemma \ref{lemma1}, we have
\begin{equation}\label{inequality}
\alpha k^3+\beta k^2+\omega k +\delta<0,
\end{equation}
where $\alpha$, $\beta$, $\omega$ and $\delta$ are defined in Theorem \ref{theorem7}. The condition (\ref{MIPERC}) is obtained by solving the cubic inequality (\ref{inequality}).
\end{proof}

\begin{rmk5}
The preliminary lemmas for Theorem \ref{theorem7}, i.e., Lemmas \ref{lemma8} and \ref{lemma1}, are tighter than the existing ones. Therefore, the MIP-based condition for ERC of BOLS in Theorem \ref{theorem7} is better than Theorem 4 in \cite{2}. This indicates a higher reconstructible sparsity level for BOLS.
\end{rmk5}

Based on Theorem \ref{theorem7}, the following lemma holds:
\begin{lemma9}
For $\mu$, $\mu_B$ and $\nu$ in Definitions \ref{def1}, \ref{def2} and \ref{def3}, suppose $(k-1)d\mu_B<1$ and $\nu=0$, then $\gamma$ in (\ref{ercdefined}) is constrained by $\gamma\leq\frac{2\mathcal{B}kd\mu_B}{2-(k-\mathcal{B})d\mu_B}<1$.
\label{lemma9}
\end{lemma9}

\subsection{Recovery Conditions in the Noisy Scenario}
Due to the noise in the actual CSS scenario, we present the performance analysis for BOLS with block-blind stopping rule under Gaussian noise, i.e., $\epsilon\sim\mathcal{N}(\mathbf{0},\sigma^2\mathbf{I}_m)$, in this subsection.
Specifically, we first use the ERC related analysis in the last subsection to further derive the theorem of exactly selecting a correct support in the current iteration for BOLS, i.e., Theorem~\ref{theorem6}. Then, the performance analysis for BOLS
with block-blind stopping rule under Gaussian noise is
investigated by developing a lower bound of the $\ell_2$-norm of the spectrum support for reliable recovery.

In the $t$-th iteration of BOLS, the residual is
\begin{equation}\label{residual}
\mathbf{r}^t=(\mathbf{I}-\mathbf{P}_{\mathbf{S}^t})\mathbf{y}=\mathbf{s}^t+\mathbf{n}^t,
\end{equation}
where $\mathbf{s}^t=(\mathbf{I}-\mathbf{P}_{\mathbf{S}^t})\mathbf{D}\mathbf{x}$ and $\mathbf{n}^t=(\mathbf{I}-\mathbf{P}_{\mathbf{S}^t})\mathbf{\epsilon}$ are called the signal and noise parts of the residual, respectively.
Denote $\gamma_{(t,1)}=||(\mathbf{D}_{0\backslash\mathbf{S}}\mathbf{R}_{0\backslash\mathbf{S}})^T\mathbf{s}^t||_{2,\infty}$, $\gamma_{(t,2)}=||(\overline{\mathbf{D}}_0\overline{\mathbf{R}}_0)^T\mathbf{s}^t||_{2,\infty}$ and $N_t=||(\mathbf{D}\mathbf{R})^T\mathbf{n}^t||_{2,\infty}$. It is known that the condition
\begin{equation}\label{condition1}
||(\mathbf{D}_{0\backslash\mathbf{S}}\mathbf{R}_{0\backslash\mathbf{S}})^T\mathbf{r}^{t}||_{2,\infty}>||(\overline{\mathbf{D}}_0\overline{\mathbf{R}}_0)^T\mathbf{r}^{t}||_{2,\infty}
\end{equation}
guarantees that the BOLS algorithm selects a correct block in the $(t+1)$-th iteration. Since $||(\mathbf{D}_{0\backslash\mathbf{S}}\mathbf{R}_{0\backslash\mathbf{S}})^T\mathbf{r}^{t}||_{2,\infty}\geq\gamma_{(t,1)}-N_t$ and $||(\overline{\mathbf{D}}_0\overline{\mathbf{R}}_0)^T\mathbf{r}^{t}||_{2,\infty}\leq\gamma_{t,2}+N_t$, we obtain that
\begin{equation}\label{suffcondition}
\gamma_{t,1}-\gamma_{t,2}>2N_t
\end{equation}
is a sufficient condition for the establishment of (\ref{condition1}). By using Lemma 4 in \cite{1} and the similar analyses in \cite{2},
\begin{equation}\label{condition2}
\gamma_{t,1}-\gamma_{t,2}\geq(1-\gamma)\gamma_{t,1}.
\end{equation}
Combining (\ref{suffcondition}), (\ref{condition2}) and Lemma \ref{lemma9}, we obtain
\begin{equation}\label{condition3}
\gamma_{(t,1)}>\frac{2(2-(k-\mathcal{B})d\mu_B)N_t}{2-(k-\mathcal{B})d\mu_B-2\mathcal{B}kd\mu_B}.
\end{equation}
In addition,
\begin{equation}\label{inadditioncondition}
\begin{aligned}
\gamma_{(t,1)}
=&||(\mathbf{Q}_{0\backslash\mathbf{S}}\mathbf{R}_{0\backslash\mathbf{S}})^T(\mathbf{I}-\mathbf{P}_{\mathbf{S}^t})\mathbf{D}_{0\backslash\mathbf{S}}\mathbf{x}_{0\backslash\mathbf{S}}||_{2,\infty}\\
\geq&\frac{||(\mathbf{Q}_{0\backslash\mathbf{S}}\mathbf{R}_{0\backslash\mathbf{S}})^T(\mathbf{I}-\mathbf{P}_{\mathbf{S}^t})\mathbf{D}_{0\backslash\mathbf{S}}\mathbf{x}_{0\backslash\mathbf{S}}||_{2}}{\sqrt{k-t}}\\
\geq&\frac{(1-(k-1)d\mu_B)||\mathbf{x}_{0\backslash\mathbf{S}}||_{2}}{\sqrt{k-t}}.
\end{aligned}
\end{equation}
The above equation and (\ref{condition3}) indicate that
\begin{equation}\label{the lastcondition}
||\mathbf{x}_{0\backslash\mathbf{S}}||_{2}>\frac{2\sqrt{k-t}(2-(k-\mathcal{B})d\mu_B)N_t}{(1-(k-1)d\mu_B)(2-(k-\mathcal{B})d\mu_B-2\mathcal{B}kd\mu_B)}
\end{equation}
guarantees the correct block selection of BOLS in the $(t+1)$-th iteration. Based on the these analyses, the following theorem holds.

\begin{theorem6}
Suppose that the condition in (\ref{MIPERC}) holds and the remaining nonzero vector $\mathbf{x}_{0\backslash\mathbf{S}}$ in the $(t+1)$-th iteration satisfies
\begin{equation}\label{BOLSselectcorrect}
\begin{aligned}
||\mathbf{x}_{0\backslash\mathbf{S}}||_2
>\frac{2\sqrt{k-t}(2-(k-\mathcal{B})d\mu_B)\sqrt{d}\sigma\sqrt{m+2\sqrt{m\log m}}}{(1-(k-1)d\mu_B)(2-(k-\mathcal{B})d\mu_B-2\mathcal{B}kd\mu_B)},
\end{aligned}
\end{equation}
then the BOLS algorithm selects one correct block in the $(t+1)$-th iteration with the probability at least $1-1/m$.
\label{theorem6}
\end{theorem6}

\begin{proof}
This theorem is proved by using (\ref{the lastcondition}) and Lemma~7 in \cite{2}.
\end{proof}

The following corollary is derived from Theorem \ref{theorem6}.
\begin{Corollary4}
Suppose that the condition in (\ref{MIPERC}) holds and all the nonzero blocks $\mathbf{x}[i]$ satisfy
\begin{equation}\label{BOLSselectcorrectcorollary}
\begin{aligned}
||\mathbf{x}[i]||_2
>\frac{2(2-(k-\mathcal{B})d\mu_B)\sqrt{d}\sigma\sqrt{m+2\sqrt{m\log m}}}{(1-(k-1)d\mu_B)(2-(k-\mathcal{B})d\mu_B-2\mathcal{B}kd\mu_B)},
\end{aligned}
\end{equation}
then the BOLS algorithm selects the true support set with the probability at least $1-1/m$.
\label{Corollary4}
\end{Corollary4}

\subsection{B-BOLS-CSS Algorithm}
In this subsection, we propose the blind stopping rule for BOLS and formulate the corresponding B-BOLS-CSS algorithm.
We denote the right hand side of (\ref{MIPERC}) as $\mathcal{C}$ and begin with the following Lemma \ref{lemma11}.

\begin{lemma11}
For Gaussian noise $\mathbf{\mathbf{\epsilon}}\sim \mathcal{N}(\mathbf{0},\sigma^2\mathbf{I}_m)$ and the measurement matrix $\mathbf{D}$ with the block-coherence $\mu_B$, we have
\begin{equation}\label{lemma91}
\begin{aligned}
{\rm Pr}\{||\mathbf{D}^T\mathbf{\epsilon}||_{2,\infty}\leq\sqrt{d}\xi\mu_B\eta\sigma\}\geq1-\frac{n}{\sqrt{2\pi\xi^2\mu_B^2\eta^2}e^{\frac{1}{2}\xi^2\mu_B^2\eta^2}},
\end{aligned}
\end{equation}
where $\xi>0$ and
\begin{equation}\label{theorem55}
\eta=\sqrt{4(m-\mathcal{C})-2}-\sqrt{m-\mathcal{C}+2\sqrt{(m-\mathcal{C})\log(m-\mathcal{C})}}.
\end{equation}
\label{lemma11}
\end{lemma11}
\begin{proof}
It is known that $||\mathbf{D}^T\mathbf{\epsilon}||_{2,\infty}\leq\sqrt{d}||\mathbf{D}^T\mathbf{\epsilon}||_{\infty}$. The lemma then follows by using Lemma 2 in \cite{4}.
\end{proof}

\begin{rmk6}
Since the variance of the element in $\mathbf{D}^T\mathbf{P}^{\bot}_{\mathbf{S}^t}\mathbf{\epsilon}$ is less than $\sigma^2$, the result in Lemma \ref{lemma11} can be extended to a new one:
\begin{equation}\label{anewone}
{\rm Pr}\{||\mathbf{D}^T\mathbf{P}^{\bot}_{\mathbf{S}^t}\mathbf{\epsilon}||_{2,\infty}\leq\sqrt{d}\xi\mu_B\eta\sigma\}>1-\frac{n}{\sqrt{2\pi\xi^2\mu_B^2\eta^2}e^{\frac{1}{2}\xi^2\mu_B^2\eta^2}}.
\end{equation}
\label{rmk6}
\end{rmk6}

\begin{theorem8}
Suppose that the condition in (\ref{MIPERC}) holds and $\mathbf{\epsilon}\sim\mathcal{N}(\mathbf{0},\sigma^2\mathbf{I}_M)$. Then, if the minimum component ${\rm SNR}_{\min}$ satisfies (\ref{snrsatismain}), with the probability
\begin{figure*}
\begin{equation}\label{snrsatismain}
\begin{aligned}
{\rm SNR}_{\min}
>\max\bigg\{&\frac{(2(2-(k-\mathcal{Y})d\mu_B)\sqrt{d}\sqrt{m+2\sqrt{m\log m}})^2}{m((1-(k-1)d\mu_B)(2-(k-\mathcal{Y})d\mu_B-2\mathcal{Y}kd\mu_B))^2},\\
&\frac{(\sqrt{d}\xi\mu_B\sqrt{m+2\sqrt{m\log m}})^2}{m(1-(k-1)d\mu_B-\sqrt{kd}\xi\mu_B(1+(k-1)d\mu_B))^2}\bigg\}
\end{aligned}
\end{equation}
\end{figure*}
\begin{equation}\label{snrprob}
\begin{aligned}
{\rm P}_{\xi}
>&1-\frac{\mathcal{C}}{m}-\frac{1}{m-\mathcal{C}}-\frac{n}{\sqrt{2\pi\xi^2\mu_B^2\eta^2}e^{\frac{1}{2}\xi^2\mu_B^2\eta^2}},
\end{aligned}
\end{equation}
the BOLS algorithm using the stopping rule
\begin{equation}\label{blindstoppingrule}
\frac{||\mathbf{D}^T\mathbf{r}^l||_{2,\infty}}{||\mathbf{r}^l||_2}\leq\sqrt{d}\xi\mu_B
\end{equation}
can reconstruct the given block $k$-sparse signal.

\label{theorem8}
\end{theorem8}

\begin{proof}
See Appendix \ref{proofoftheorem8}.
\end{proof}

By further utilizing Lemma \ref{lemma11}, we obtain the following corollary, which provides a tighter bound for ${\rm SNR}_{\min}$.

\begin{Corollary1}
Suppose that the condition in (\ref{MIPERC}) holds and $\mathbf{\epsilon}\sim\mathcal{N}(\mathbf{0},\sigma^2\mathbf{I}_M)$. Then, if the minimum component ${\rm SNR}_{\min}$ satisfies (\ref{snrsatismain2}), with the probability
\begin{figure*}
\begin{equation}\label{snrsatismain2}
\begin{aligned}
{\rm SNR}_{\min}
>\max\bigg\{&\frac{(2(2-(k-\mathcal{Y})d\mu_B)\sqrt{d}\xi\mu_B\eta)^2}{m((1-(k-1)d\mu_B)(2-(k-\mathcal{Y})d\mu_B-2\mathcal{Y}kd\mu_B))^2},\\
&\frac{(\sqrt{d}\xi\mu_B\sqrt{m+2\sqrt{m\log m}})^2}{m(1-(k-1)d\mu_B-\sqrt{kd}\xi\mu_B(1+(k-1)d\mu_B))^2}\bigg\}
\end{aligned}
\end{equation}
\end{figure*}
\begin{equation}\label{snrprob2}
\begin{aligned}
{\rm P}_{\xi}
>&1-\frac{1}{m}-\frac{1}{m-\mathcal{C}}-\frac{\mathcal{C}n}{\sqrt{2\pi\xi^2\mu_B^2\eta^2}e^{\frac{1}{2}\xi^2\mu_B^2\eta^2}},
\end{aligned}
\end{equation}
the BOLS algorithm using the stopping rule (\ref{blindstoppingrule})
can reconstruct the given block $k$-sparse signal,
where $\xi>0$ and $\eta$ is provided in (\ref{theorem55}).

\label{corollary1}
\end{Corollary1}

The rule in (\ref{blindstoppingrule}) is called the blind stopping rule for BOLS, which does not depend on the sparsity level or the noise information. The $\ell_{2,\infty}$ term in (\ref{blindstoppingrule}) deals with the signal in blocks, which can be regarded as the energy sum of the outputs of multiple matched filters for one signal block. In this way, the stopping rule can blindly decide whether there exist matrix blocks corresponding to support blocks of the sparse spectrum in the remaining measurement matrix. The probability in (\ref{snrprob}) is actually the target probability of detection. With a given ${\rm P}_{\xi}$, $\xi$ can be calculated according to (\ref{snrprob}).
Based on these descriptions, the B-BOLS-CSS algorithm is given in Algorithm \ref{alg:11}.

\begin{algorithm}
	\renewcommand{\algorithmicrequire}{\textbf{Input:}}
	\renewcommand{\algorithmicensure}{\textbf{Output:}}
	\caption{B-BOLS-CSS Algorithm}
	\label{alg:11}
	\begin{algorithmic}[1]
		\REQUIRE $\mathbf{D}, \mathbf{y}$, detection probability ${\rm P}_{\xi}$ and the block length $d$
		\ENSURE The recovered spectrum $\hat{\mathbf{x}}\in\mathcal{R}^{N}$ and the set of the support indices $\hat{\mathbf{S}}\subseteq \{1,2,\cdots,N\}$
        \STATE $\mathbf{Initialization:}$ $t=0$, $\mathbf{r}^0=\mathbf{y}$, $\mathbf{S}^0=\emptyset$, $\mathbf{x}^0=\mathbf{0}$
        \STATE Calculate $\xi$ according to (\ref{snrprob}) with the given ${\rm P}$
        \STATE Calculate the block-coherence $\mu_B$ of $\mathbf{D}$
        \WHILE {$\frac{||\mathbf{D}^T\mathbf{r}^t||_{2,\infty}}{||\mathbf{r}^t||_2}>\sqrt{d}\xi\mu_B$}
		\STATE Set $i^{t+1}=\mathop{\arg\min}\limits_{j\in\{1,\cdots,N\}\backslash\mathbf{S}^{t}}||\mathbf{P}^\bot_{\mathbf{S}^{t}\cup \{(j-1)d:jd\}}\mathbf{y}||_2^2$
		\STATE Augment $\mathbf{S}^{t+1}=\mathbf{S}^{t}\cup{\{i^{t+1}\}}$
		\STATE Estimate $\mathbf{x}^{t+1}=\mathop{\arg\min}\limits_{\mathbf{x}: \;{\rm supp}(\mathbf{x})=\mathbf{S}^{t+1}}\|\mathbf{y}-\mathbf{D}\mathbf{x}\|_2^2$
        \STATE Update $\mathbf{r}^{t+1}=\mathbf{y}-\mathbf{D}\mathbf{x}^{t+1}$
		\STATE $t=t+1$
        \ENDWHILE
		\STATE \textbf{return} $\hat{\mathbf{S}}=\mathbf{S}^{t}$ and $\hat{\mathbf{x}}=\mathbf{x}^{t}$
	\end{algorithmic}
\end{algorithm}

\section{Simulation Results}
\label{simulation}

In this section, we first perform simulations to illustrate our theoretical results presented in Section III and compare them with existing ones. Then, we compare our proposed B-BOLS-CSS algorithm with the state-of-art ones.

\subsection{Simulations for Theoretical Results}

\subsubsection{Comparison between Lemma \ref{lemma8} and Lemma 2 in \cite{1}}
\begin{figure} \centering 
	\subfigure[] { \label{fig:a} 
		\includegraphics[width=0.466\columnwidth]{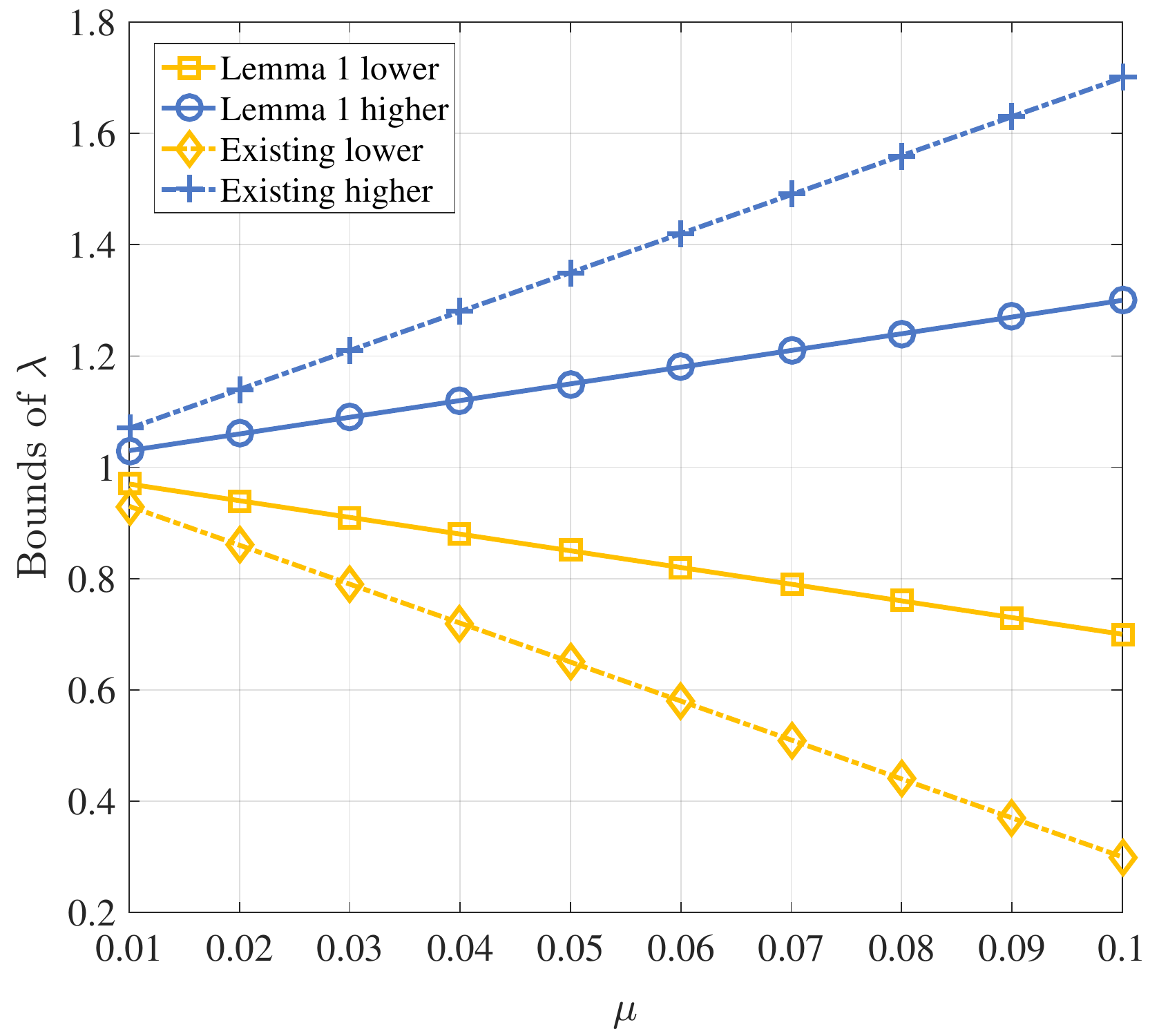} 
	} 
	\subfigure[] { \label{fig:b} 
		\includegraphics[width=0.466\columnwidth]{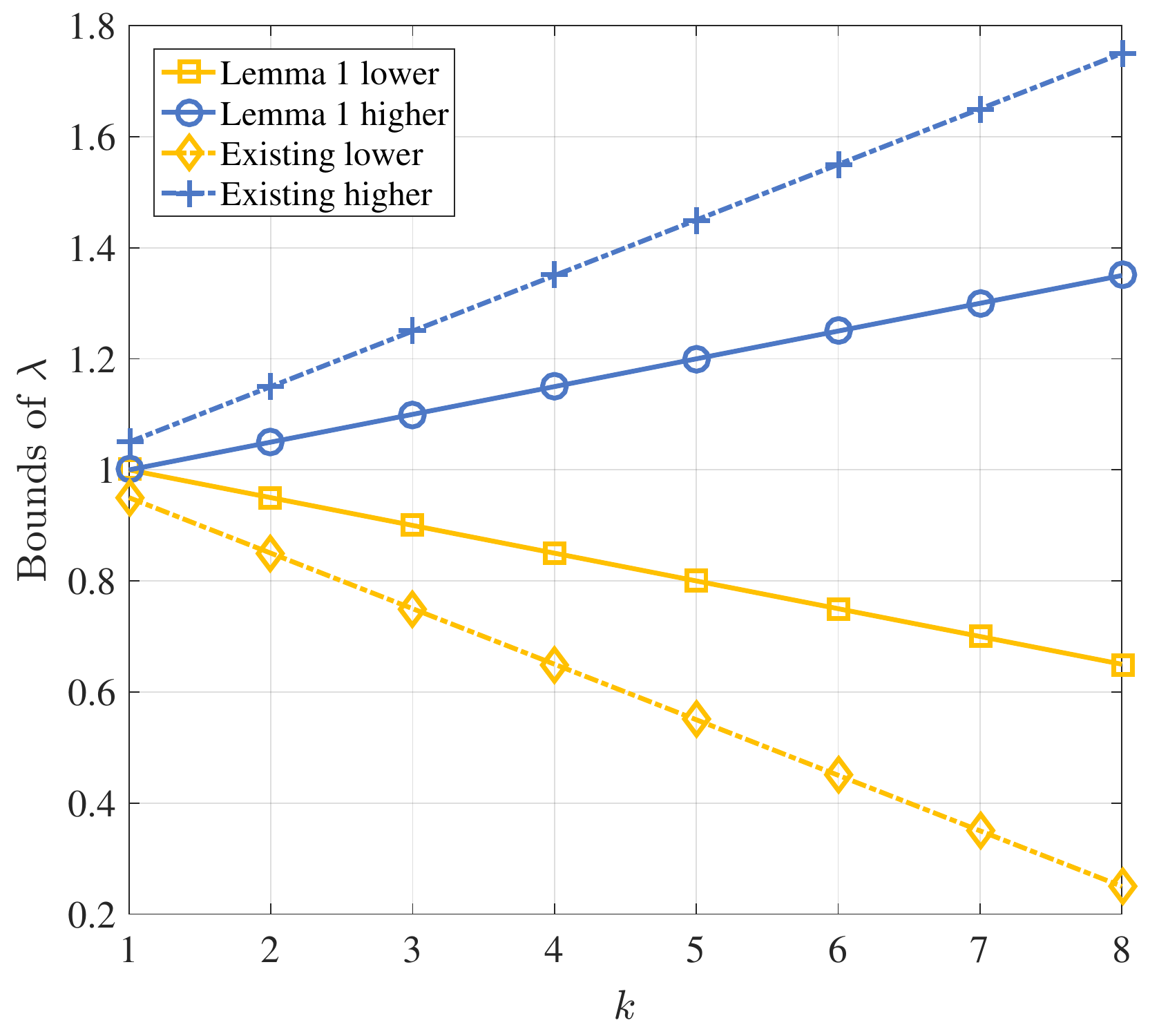} 
	} 
	\caption{The bounds of $\lambda$ versus (a) $\mu$ with $\mu_B=\frac{\mu}{d}$, $k=4$ and $d=2$; (b) $k$ with $\mu=0.05$, $\mu_B=\frac{\mu}{d}$ and $d=2$.}
	\label{lambdabounds1} 
\end{figure}

The lower bound and the higher bound in Lemma \ref{lemma8} are called ``Lemma 1 lower'' and ``Lemma 1 higher'' respectively. Accordingly, the lower bound and the higher bound in Lemma 2 of \cite{1} are called ``Existing lower'' and ``Existing higher''.
As shown in Fig. \ref{lambdabounds1}, our derived bounds are tighter than the existing ones, i.e., they are much closer to 1. With the increase of matrix coherence $\mu$ or the block sparsity $k$, the bounds become away from 1 but our derived ones still keep considerable tightness, which indicates that the follow-up theoretical analysis based on Lemma
\ref{lemma8} may provide better result.

\subsubsection{Comparison among Lemma \ref{lemma1}, the existing bounds in (\ref{existingbound1}) and (\ref{existingbound2})}
\begin{figure} \centering 
	\subfigure[] { \label{fig:a} 
		\includegraphics[width=0.466\columnwidth]{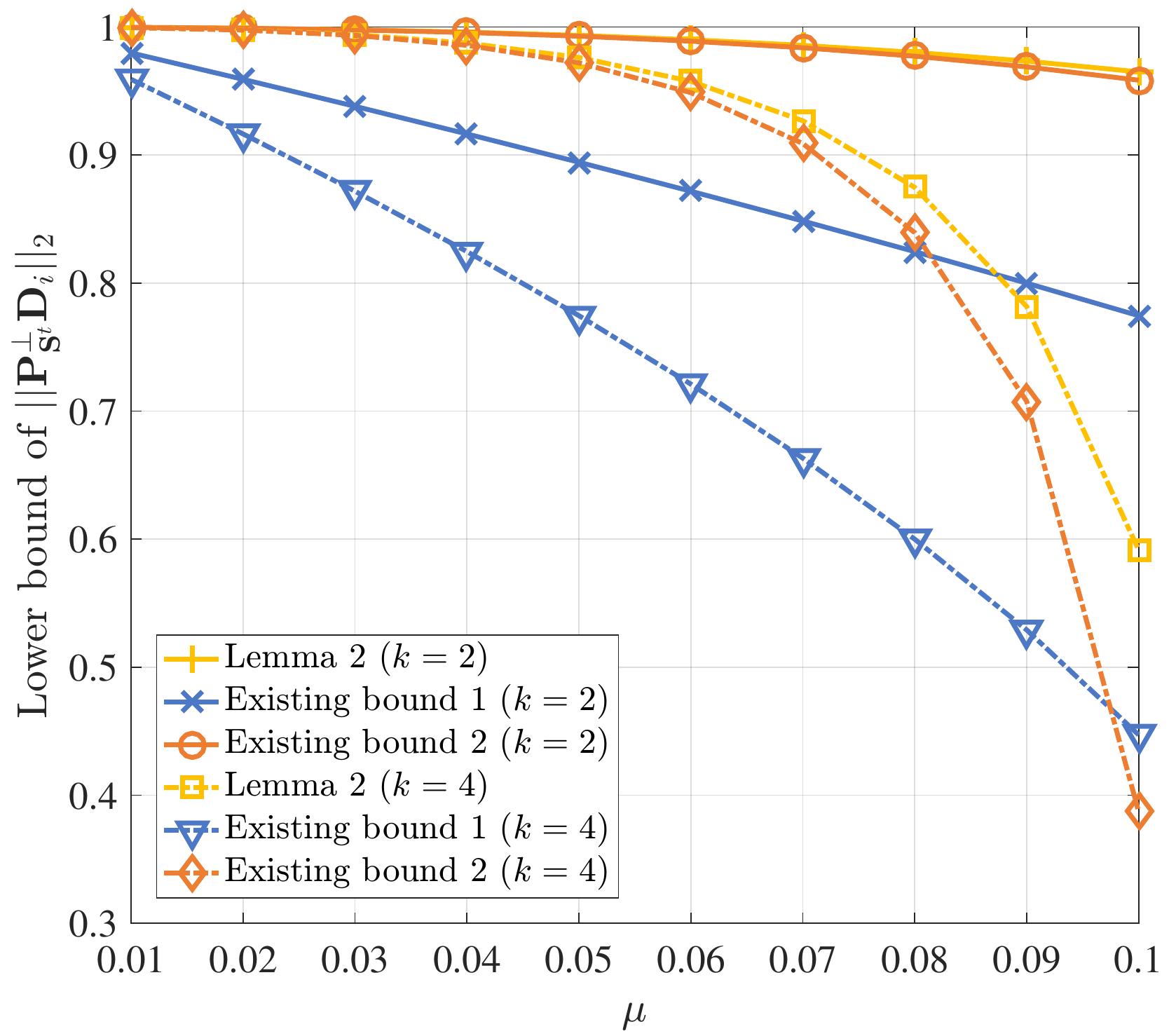} 
	} 
	\subfigure[] { \label{fig:b} 
		\includegraphics[width=0.466\columnwidth]{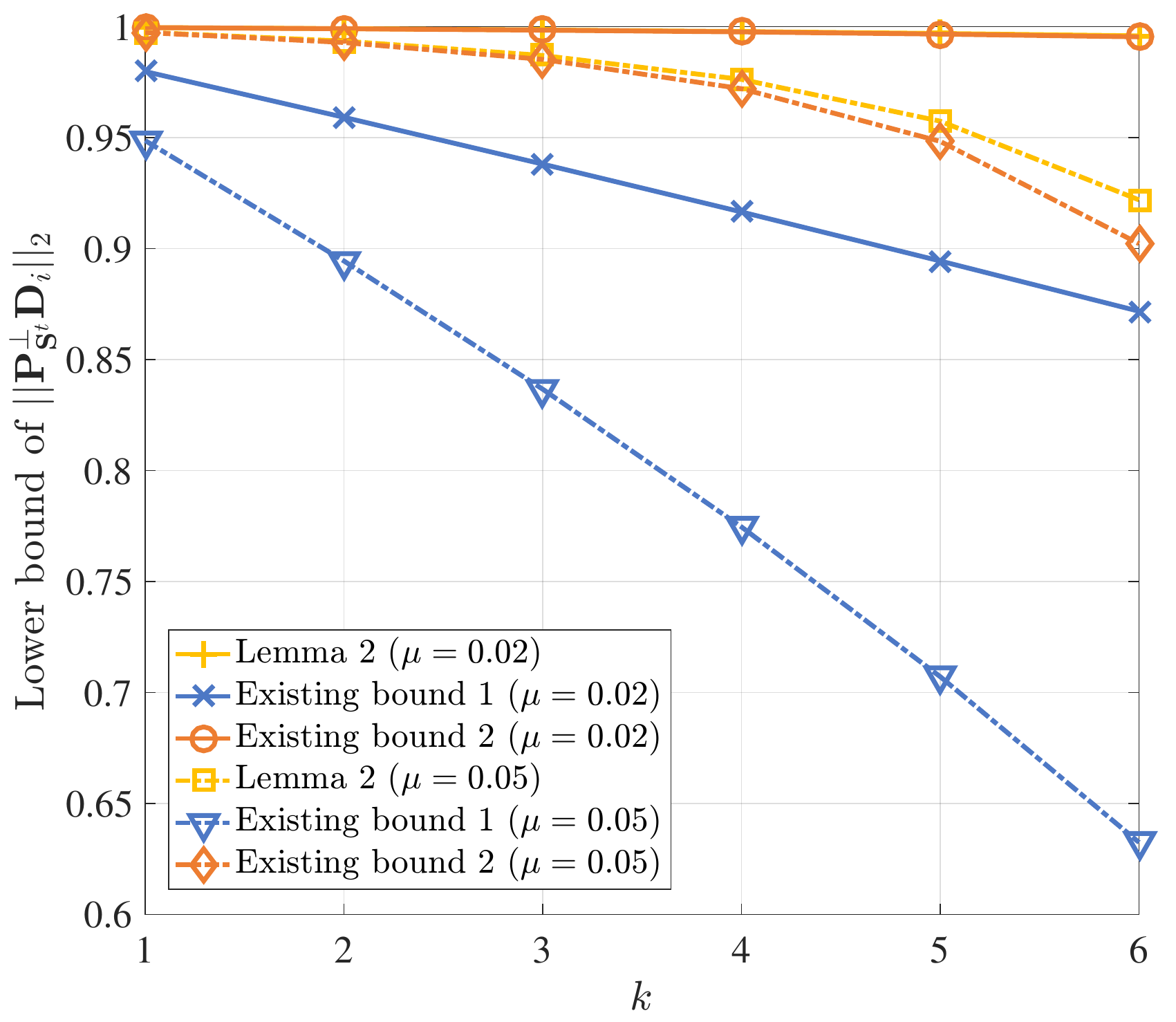} 
	} 
	\caption{The lower bounds of $||\mathbf{P}_{\mathbf{S}^t}^{\bot}\mathbf{D}_i||_2$ with $\mu_B=\frac{\mu}{d}$ and $d=2$.}
	\label{boundps1} 
\end{figure}

The existing bounds in (\ref{existingbound1}) and (\ref{existingbound2}) are called ``Existing bound 1'' and ``Existing bound 2'' respectively. As illustrated in Fig. \ref{boundps1}, our derived result in Lemma \ref{lemma1} is always much closer to 1, which indicates that this result is tighter than the existing ones. Although a larger $\mu$ or a larger $k$ causes degradation of the bound, the result in Lemma \ref{lemma1} performs the best.

\subsubsection{Comparison between Theorem \ref{theorem7} and Theorem 4 in \cite{2}}

\begin{figure}
  \centering
  \includegraphics[scale=0.43]{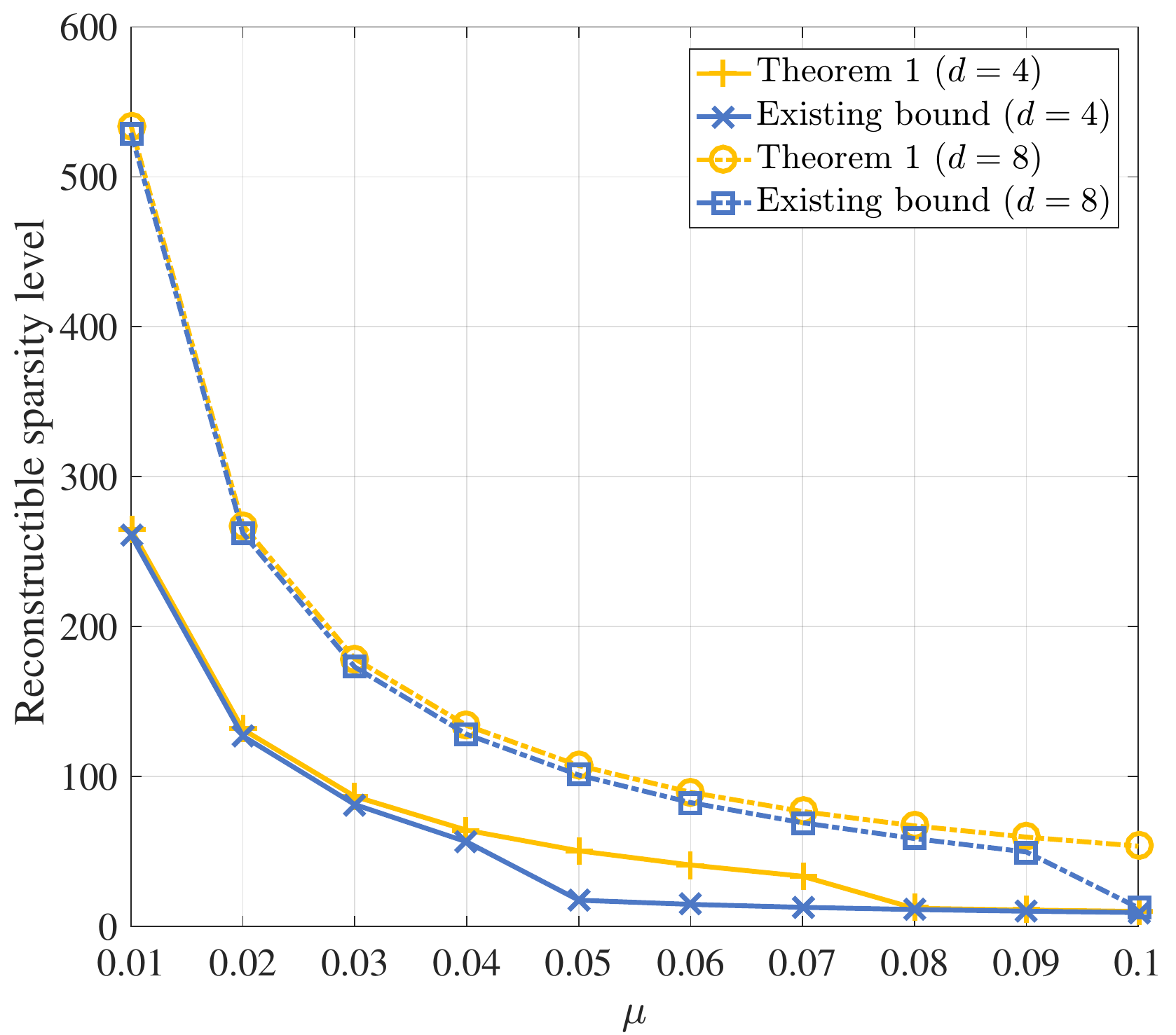}\\
  \caption{The bounds of reconstructible sparsity level versus $\mu$ with $\mu_B=\frac{\mu}{d}$.}\label{sparsitylevel}
\end{figure}

The result of Theorem 4 in \cite{2} is called ``Existing bound''. As presented in Fig. \ref{sparsitylevel}, our derived result of the reconstructible sparsity level is higher than the existing one. This indicates that our result presents a more relaxed theoretical reconstructible sparsity level for BOLS algorithm. Furthermore, this improved theoretical analysis lays a beneficial foundation for the subsequent analysis of the noisy recovery performance and the blind stopping mechanism for BOLS.

\subsubsection{Comparison of the lower bounds of ${\rm SNR}_{\min}$ in Theorem \ref{theorem8} and the ${\rm SNR}_{\min}$ in Theorem 1 of \cite{4}}

The lower bound of ${\rm SNR}_{\min}$ in Theorem 1 of \cite{4} is called ``Existing bound''. The dimensions of the measurement matrix are set as $m=1024$, $n=8192$ and $m=2048$, $n=8192$, which are the same as those in the simulations of \cite{4}. The corresponding coherence $\mu$ are 0.135 and 0.109 respectively. The probabilities of exact recovery ${\rm P}$ in (\ref{snrprob}) are fixed to $0.9:0.01:0.99$.

\begin{figure}
  \centering
  \includegraphics[scale=0.43]{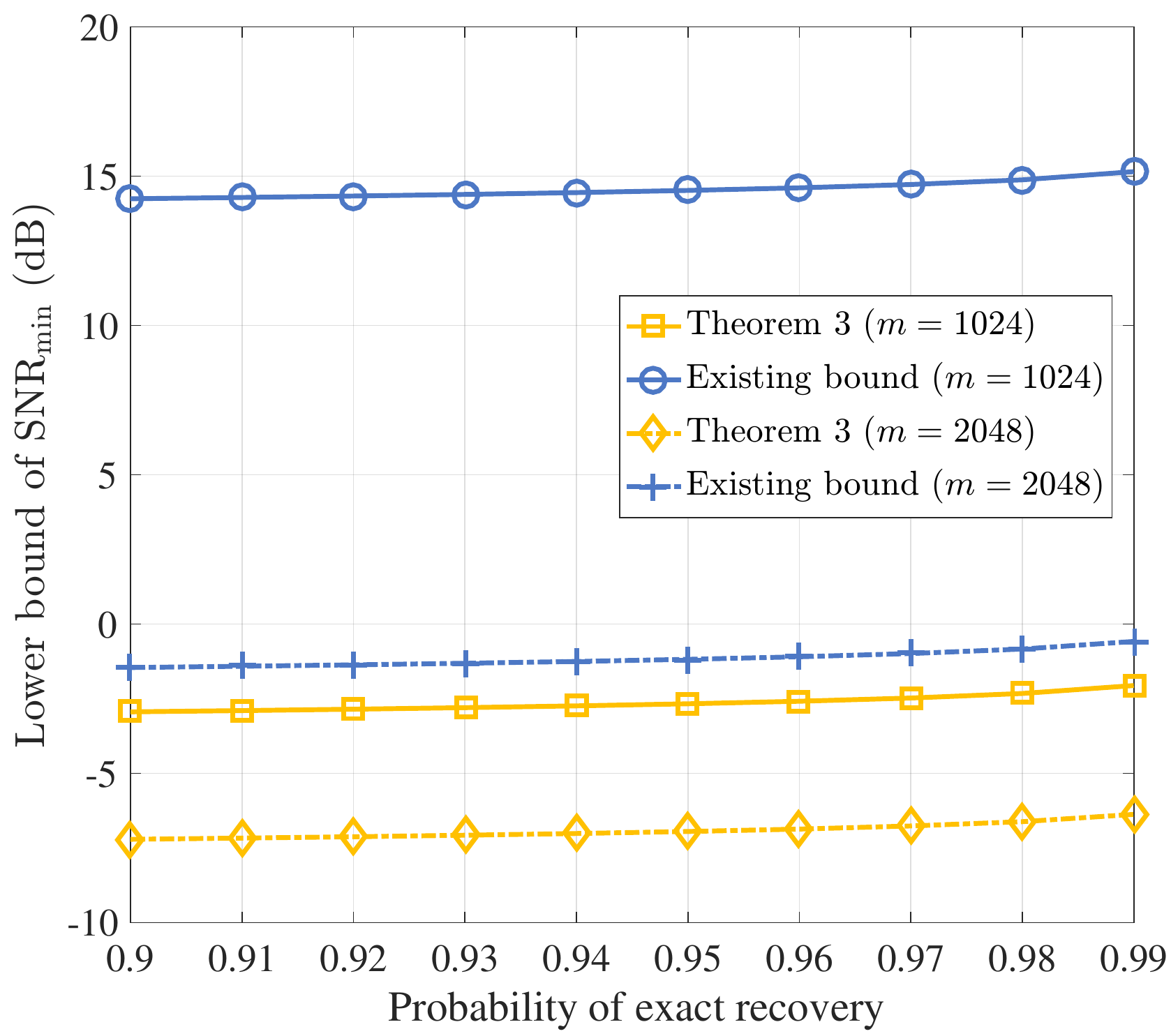}\\
  \caption{The lower bounds of ${\rm SNR}_{\min}$ versus probability of exact recovery with $\mu_B=\frac{\mu}{d}$, $k=2$, $d=2$ and $n=8192$.}\label{snrcom}
\end{figure}

As shown in Fig. \ref{snrcom}, the lower bounds of ${\rm SNR}_{\min}$ given in (\ref{snrsatismain}) are lower than those in \cite{4}, which indicates that BOLS performs better than OMP under low SNR condition. Meanwhile, BOLS is more capable to achieve the target probability of exact recovery even if the number of measurements of BOLS is only half of that of OMP, leading to computing resource savings.

\subsection{Simulations for CSS}
In this subsection, we perform simulations to compare our proposed B-BOLS-CSS algorithms with the other CSS algorithms.

Consider two wideband CR systems with $m=128$, $n=512$ and $m=256$, $n=512$. The locations of the nonzero blocks of the sparse spectrum are selected uniformly at random. Two types of the measurement matrices are generated. The first type is the widely used Gaussian measurement matrix and the elements in this matrix are independently and identically distributed as $\mathcal{N}(0,1/m)$ with block orthogonality, i.e., $\nu=0$.
The second one is the hybrid measurement matrix, which is used in \cite{14}. The column of the measurement matrix is set as $\mathbf{D}_i=a_i(\mathbf{h}_i+g_i\mathbf{1})$, where $\mathbf{h}_i$ satisfies the standard Gaussian distribution, $\mathbf{1}$ is the all 1 vector and $g_i$ obeys the uniform distribution on $[0,G]$ with $G>0$. Note that the MIP of a hybrid measurement matrix is bad, which is close to 1 and is much higher than that of a Gaussian measurement matrix. All realizations of the measurement matrix are normalized. The recovery is successful if the recovered spectrum vector is within a certain small Euclidean distance of the original spectrum. For each trial, we average over 1,000 realizations of the sparse spectrum.

\subsubsection{CSS Using Gaussian Measurement Matrix}

The nonzero entries of the spectrum are independently and identically distributed as $\mathcal{N}(0,1)$.
The CSS schemes using OLS, BOLS, OMP and BOMP are called ``OLS-CSS'', ``BOLS-CSS'', ``OMP-CSS'' and ``BOMP-CSS'', all iterating for exact $k$ or $kd$ times. The CSS using the OMP with blind stopping rule in \cite{4} is called ``B-OMP-CSS''.

\begin{figure}
  \centering
  \includegraphics[scale=0.43]{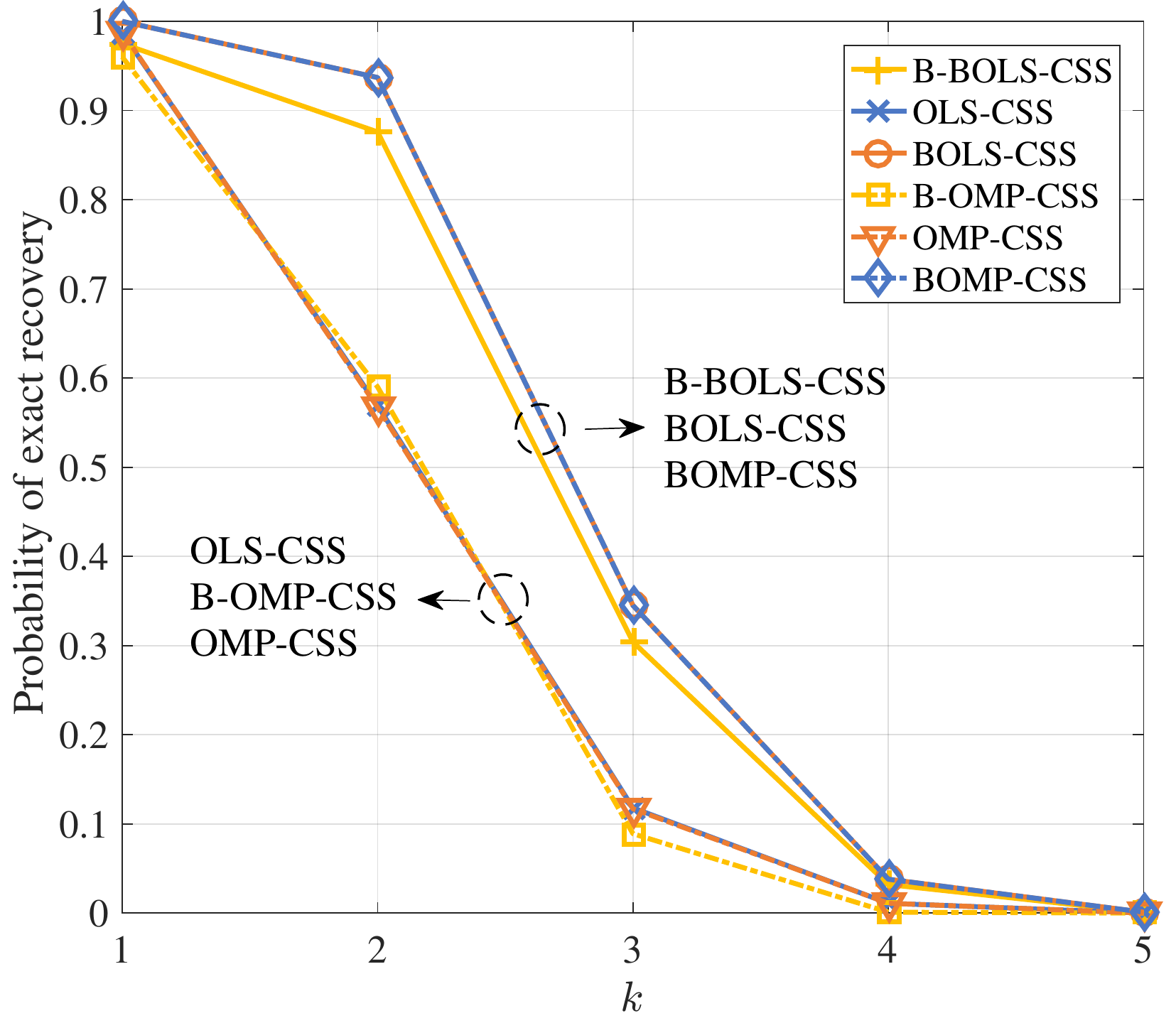}\\
  \caption{Probabilities of exact recovery versus $k$ using Gaussian measurement matrix with $m=128$, $n=512$ and $d=4$.}\label{k128512}
\end{figure}

\begin{figure}
  \centering
  \includegraphics[scale=0.43]{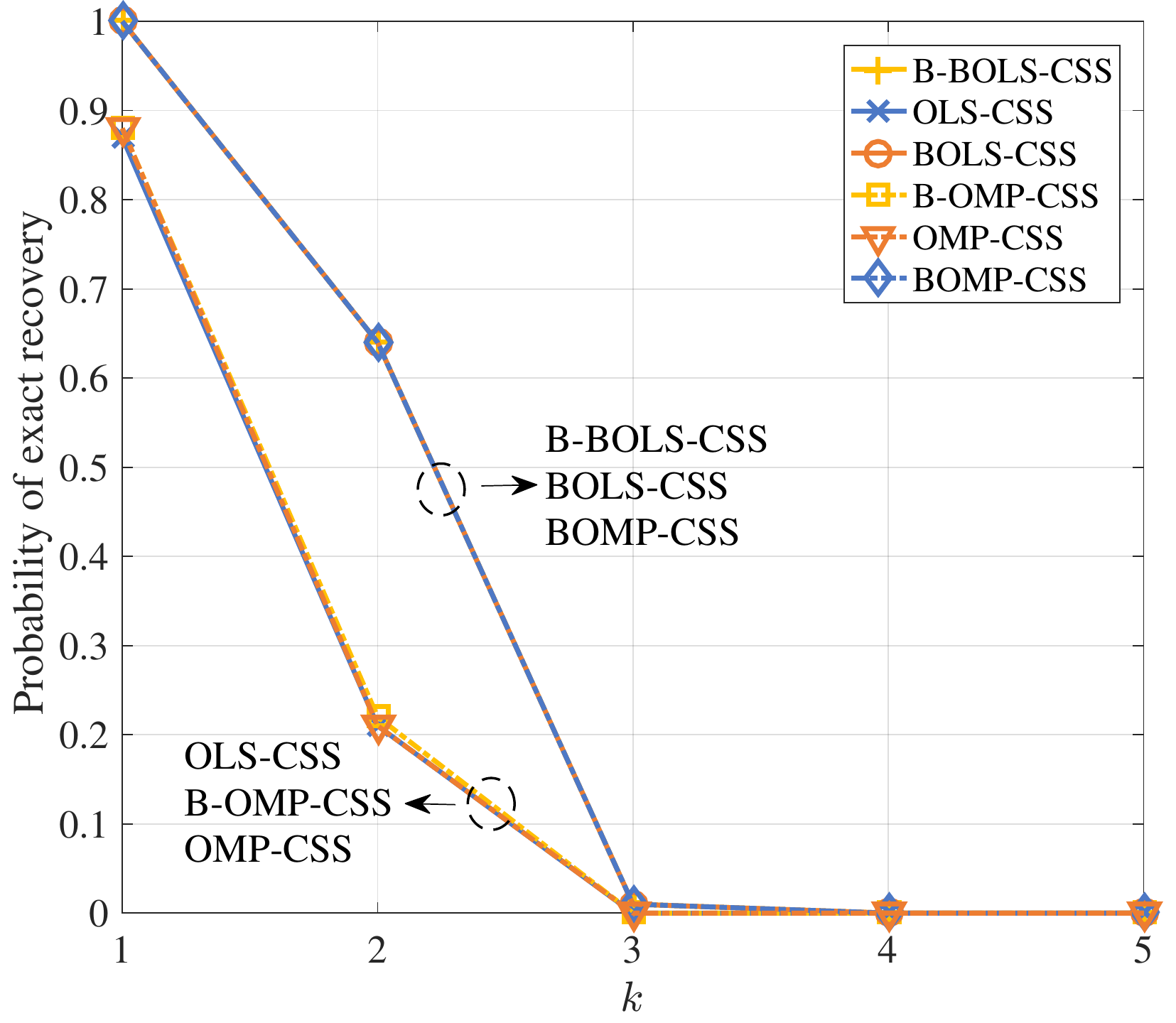}\\
  \caption{Probabilities of exact recovery versus $k$ using Gaussian measurement matrix with $m=256$, $n=512$ and $d=8$.}\label{k256512}
\end{figure}

In Figs. \ref{k128512} and \ref{k256512}, we plot the probability of exact recovery as a function of the block sparsity $k$ of the generated spectrum where the SNR is fixed as 20 dB. The number of measurements of the measurement matrices are fixed as $m=128$ and $m=256$, respectively. It is observed that the CSS algorithms using block structure outperform the ones without employing block characteristics. The sensing performance of our proposed B-BOLS-CSS is comparable with that of BOLS-CSS and BOMP-CSS, which indicates that B-BOLS-CSS iterates for appropriate times and thus effectively deals with unknown prior information issue. Meanwhile, the performance of B-BOLS-CSS is better than that of B-OMP-CSS, which reveals that the utilization of block property is an effective way in improving the accuracy and B-BOLS-CSS is really an attractive choice for wideband spectrum sensing.

\begin{figure}
  \centering
  \includegraphics[scale=0.43]{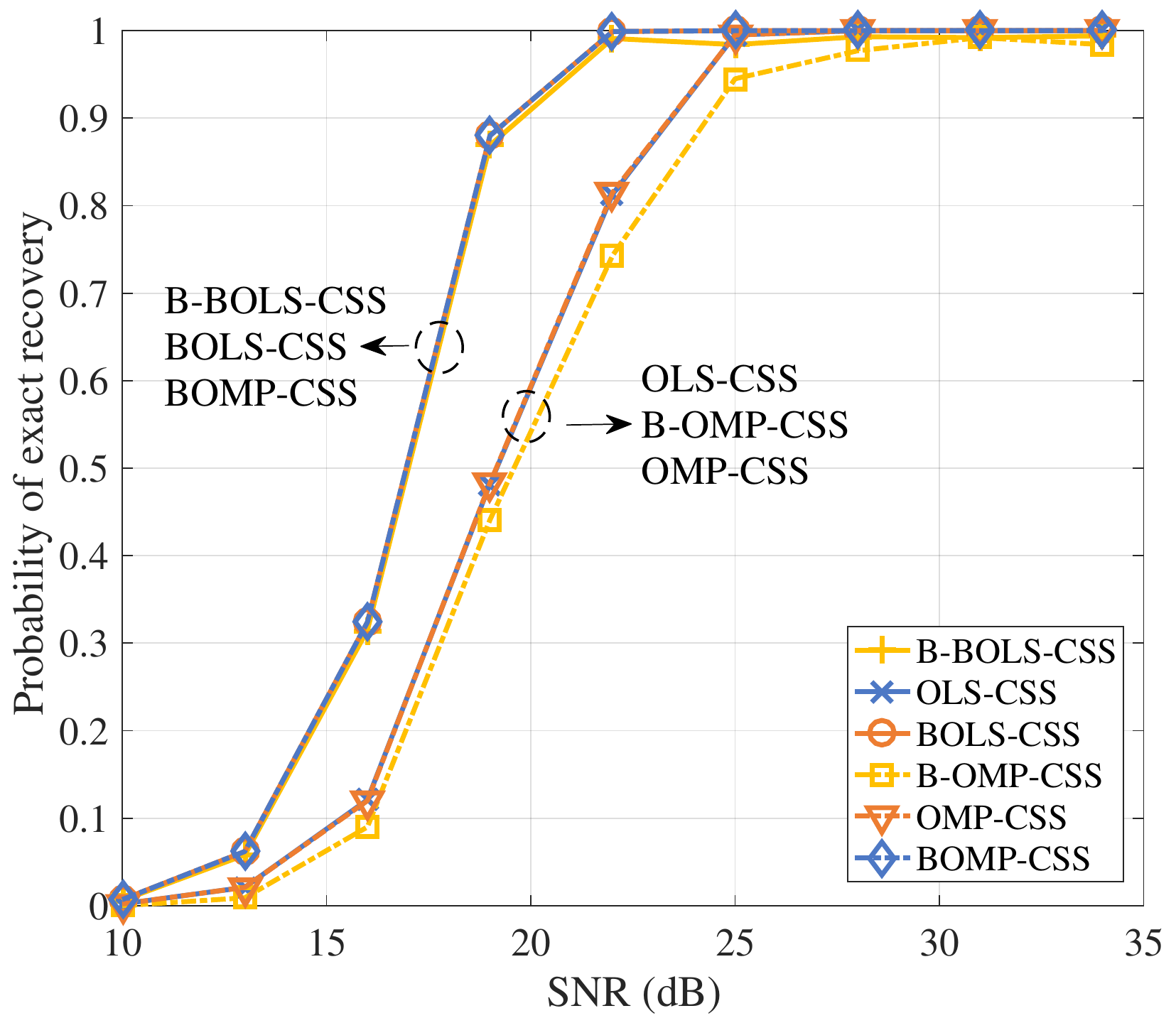}\\
  \caption{Probabilities of exact recovery versus SNR (dB) using Gaussian measurement matrix with $m=128$, $n=512$ and $d=4$.}\label{snr128512}
\end{figure}

\begin{figure}
  \centering
  \includegraphics[scale=0.43]{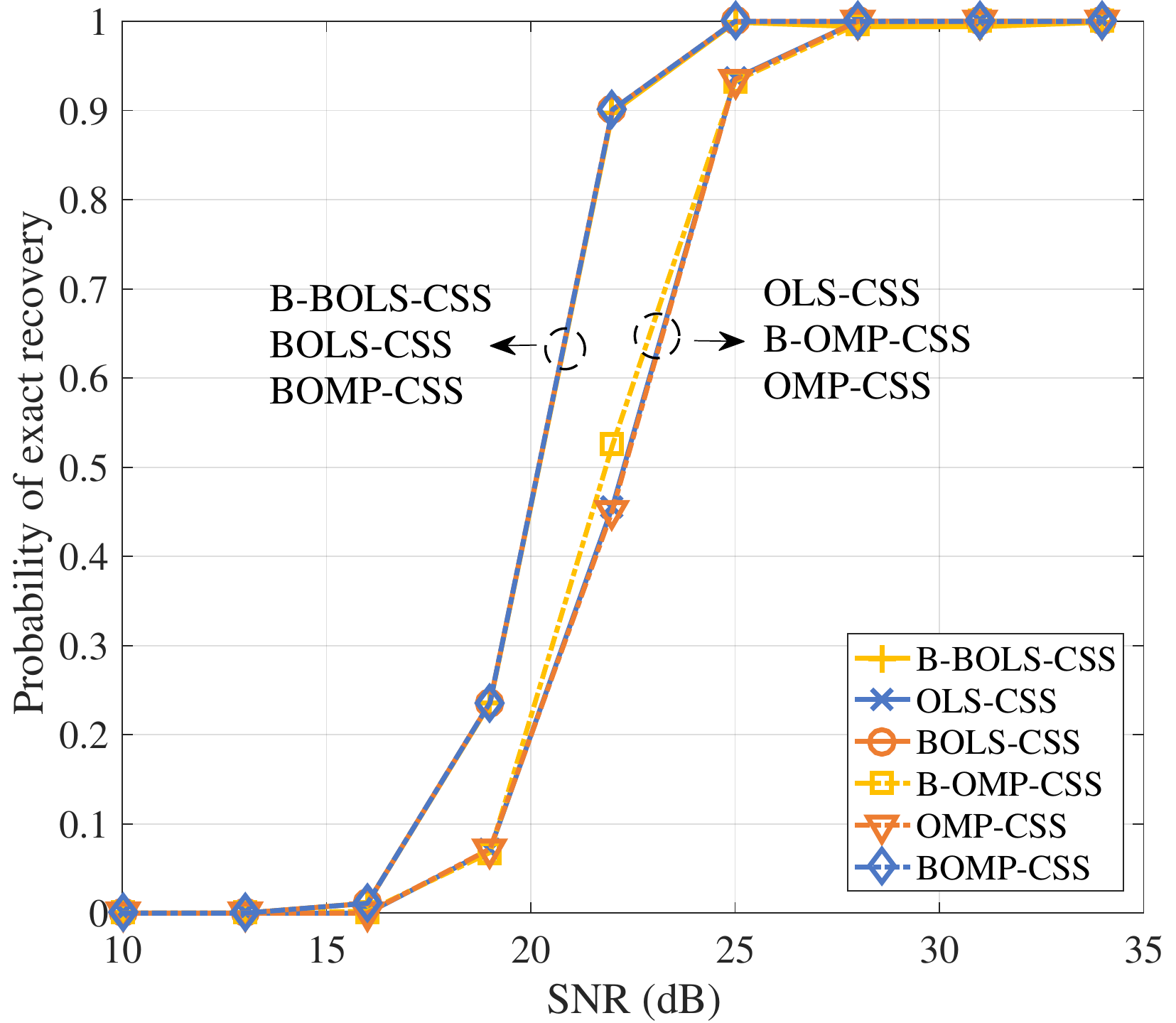}\\
  \caption{Probabilities of exact recovery versus SNR (dB) using Gaussian measurement matrix with $m=256$, $n=512$ and $d=8$.}\label{snr256512}
\end{figure}

In Figs. \ref{snr128512} and \ref{snr256512}, we plot the probability of exact recovery when SNR (dB) varies. It can be seen that the performance of the block CSS algorithms is better than those of the conventional CSS algorithms in both low and high SNR scenarios.
Meanwhile, B-BOLS-CSS can still exhibit its reliable recovery performance, which indicates its robustness in CSS. When compared with Figs. \ref{k128512} and \ref{k256512}, the similar conclusions can be obtained in Figs. \ref{snr128512} and \ref{snr256512}.

\subsubsection{CSS Using Hybrid Measurement Matrix}
The nonzero atoms of the spectrum are independently and identically
distributed as either $\mathcal{N}(0,1)$ or $\mathcal{N}(1,0.01)$ to further see the influence of different amplitude distributions on the sensing performance.



\begin{figure}
  \centering
  \includegraphics[scale=0.43]{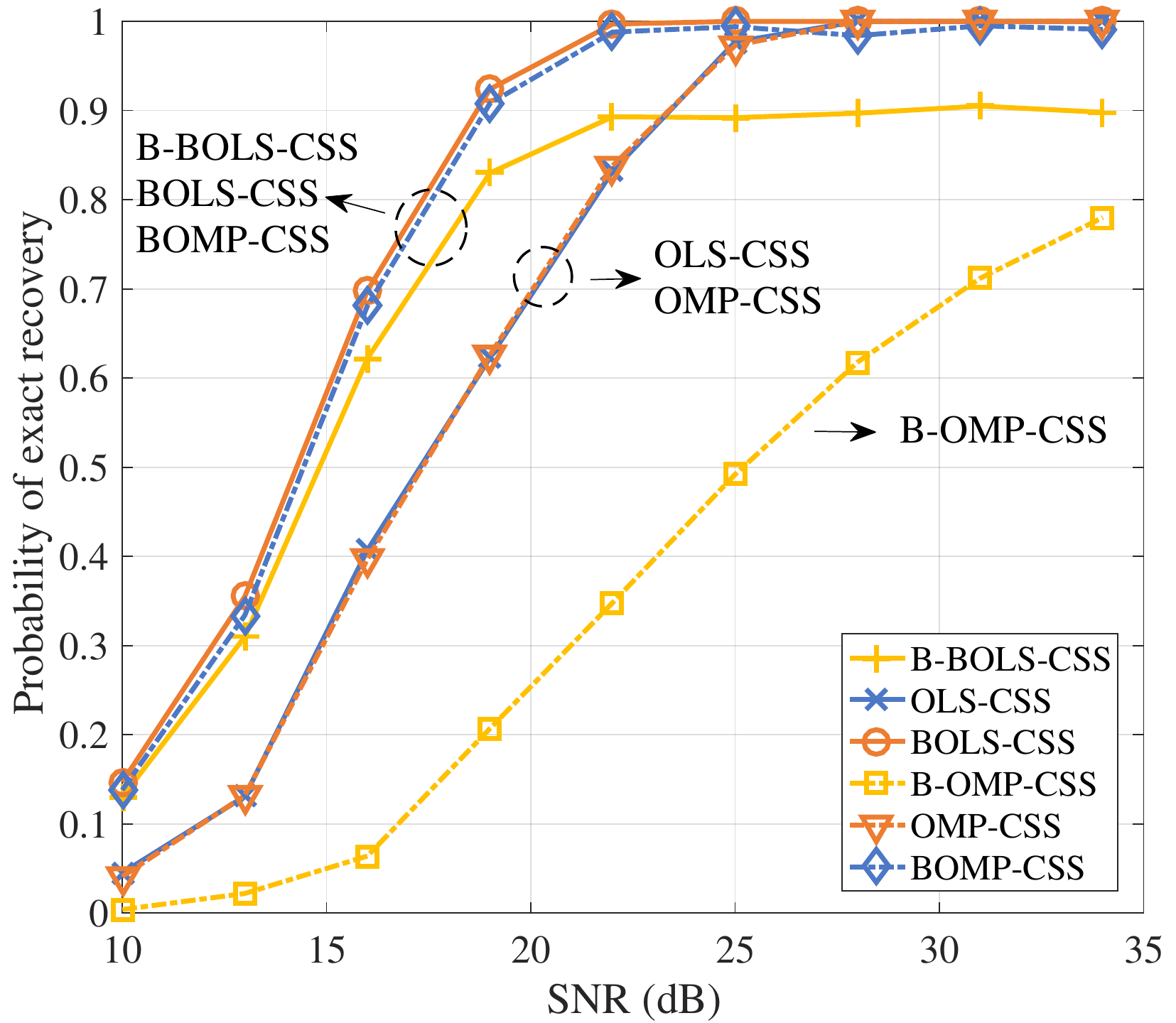}\\
  \caption{Probabilities of exact recovery versus SNR (dB) using hybrid measurement matrix with $m=128$, $n=512$, $d=4$ and the entries of the sparse spectrum satisfy $\mathcal{N}(0,1)$.}\label{measurement2snr128512d4}
\end{figure}

In Fig. \ref{measurement2snr128512d4}, the probabilities of exact recovery of BOLS-CSS are better than those of BOMP-CSS. It is known that the hybrid measurement matrix expresses a particularly poor MIP which is close to 1. It thus reveals that BOLS is more suitable for wideband CSS since BOLS-CSS exhibits reliable performance even if the MIP of the measurement matrix is unsatisfactory. Meanwhile, for the blind CSS algorithm, B-BOLS-CSS is always better than B-OMP-CSS. This indicates that the atomic selection mechanism in BOLS and the consideration of block sparsity are indeed conducive to improve CSS performance.

\begin{figure}
  \centering
  \includegraphics[scale=0.43]{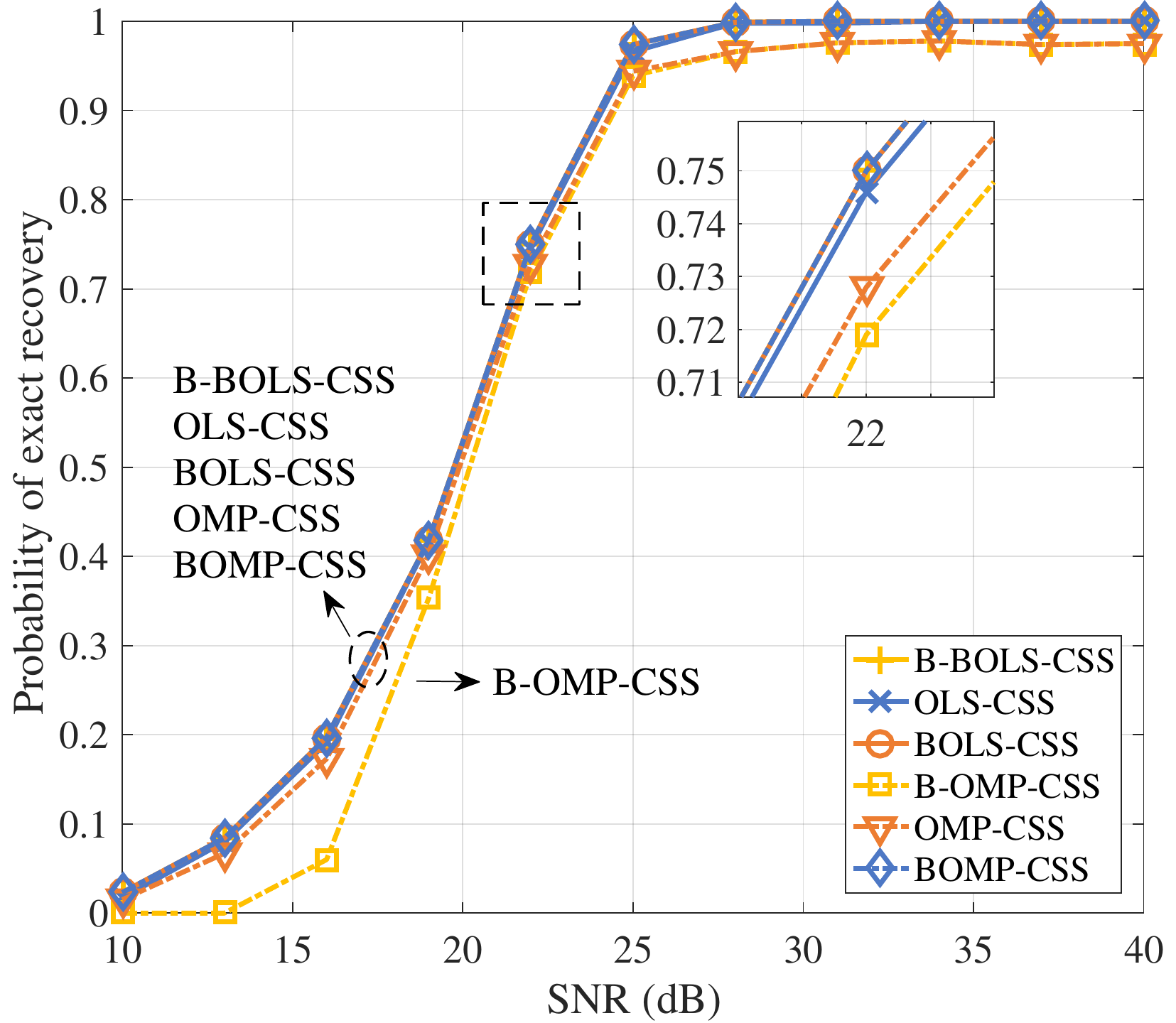}\\
  \caption{Probabilities of exact recovery versus SNR (dB) using hybrid measurement matrix with $m=128$, $n=512$, $d=4$ and the entries of the sparse spectrum satisfy $\mathcal{N}(1,0.01)$.}\label{measurement2snr128512d4model2}
\end{figure}

In Fig. \ref{measurement2snr128512d4model2}, the sensing performance of the OLS-type algorithms is better than that of the OMP-type algorithms, which reveals that OLS-type algorithms are able to deal with different distributions of the spectrum support. Meanwhile, the sensing performance of B-BOLS-CSS is competitive with that of the BOLS-CSS, indicating the effectiveness and robustness of B-BOLS-CSS when prior information is absent.

\begin{figure}
  \centering
  \includegraphics[scale=0.43]{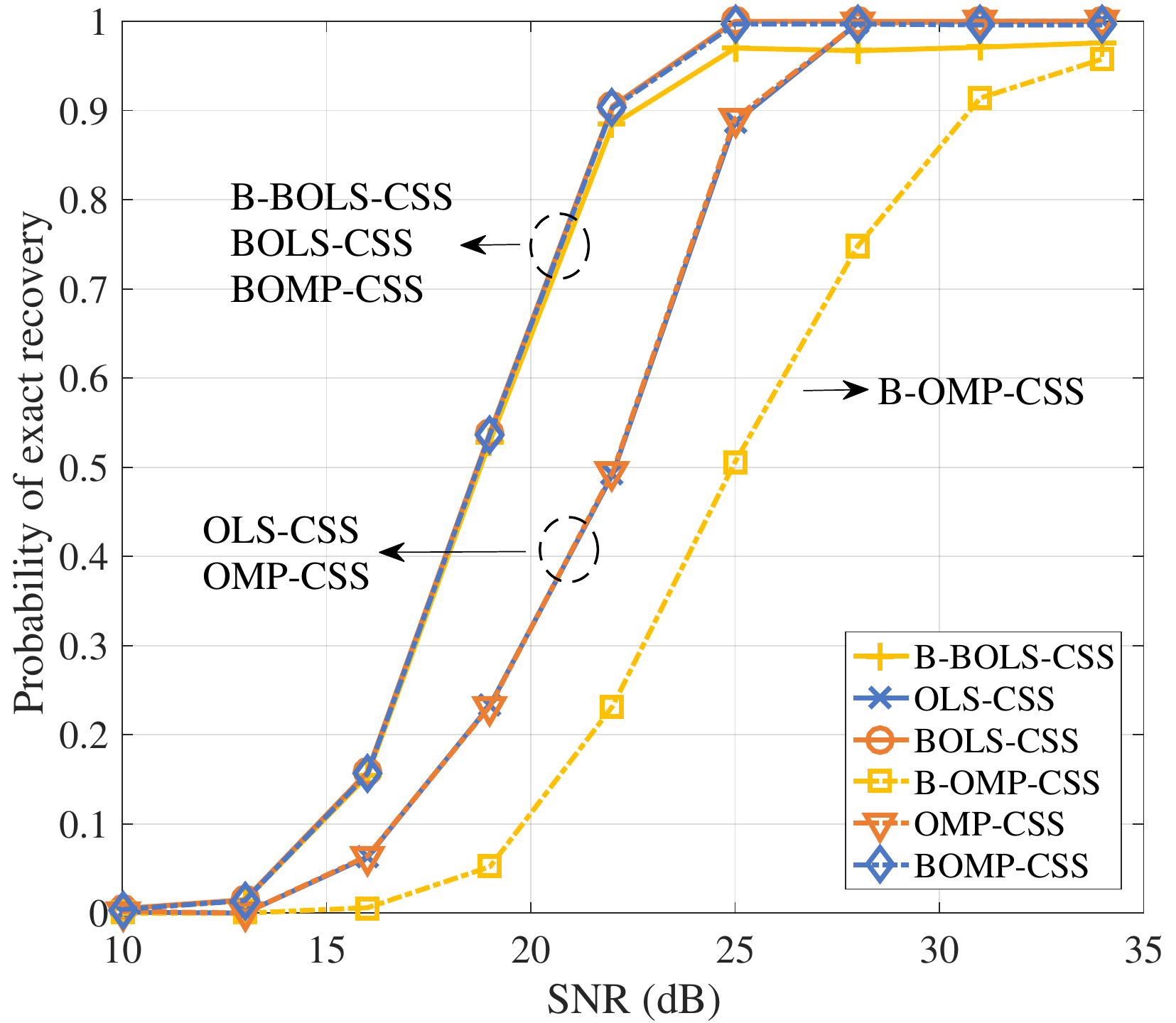}\\
  \caption{Probabilities of exact recovery versus SNR (dB) using hybrid measurement matrix with $m=256$, $n=512$, $d=8$ and the entries of the sparse spectrum satisfy $\mathcal{N}(0,1)$.}\label{measurement2snr256512d8}
\end{figure}

\begin{figure}
  \centering
  \includegraphics[scale=0.43]{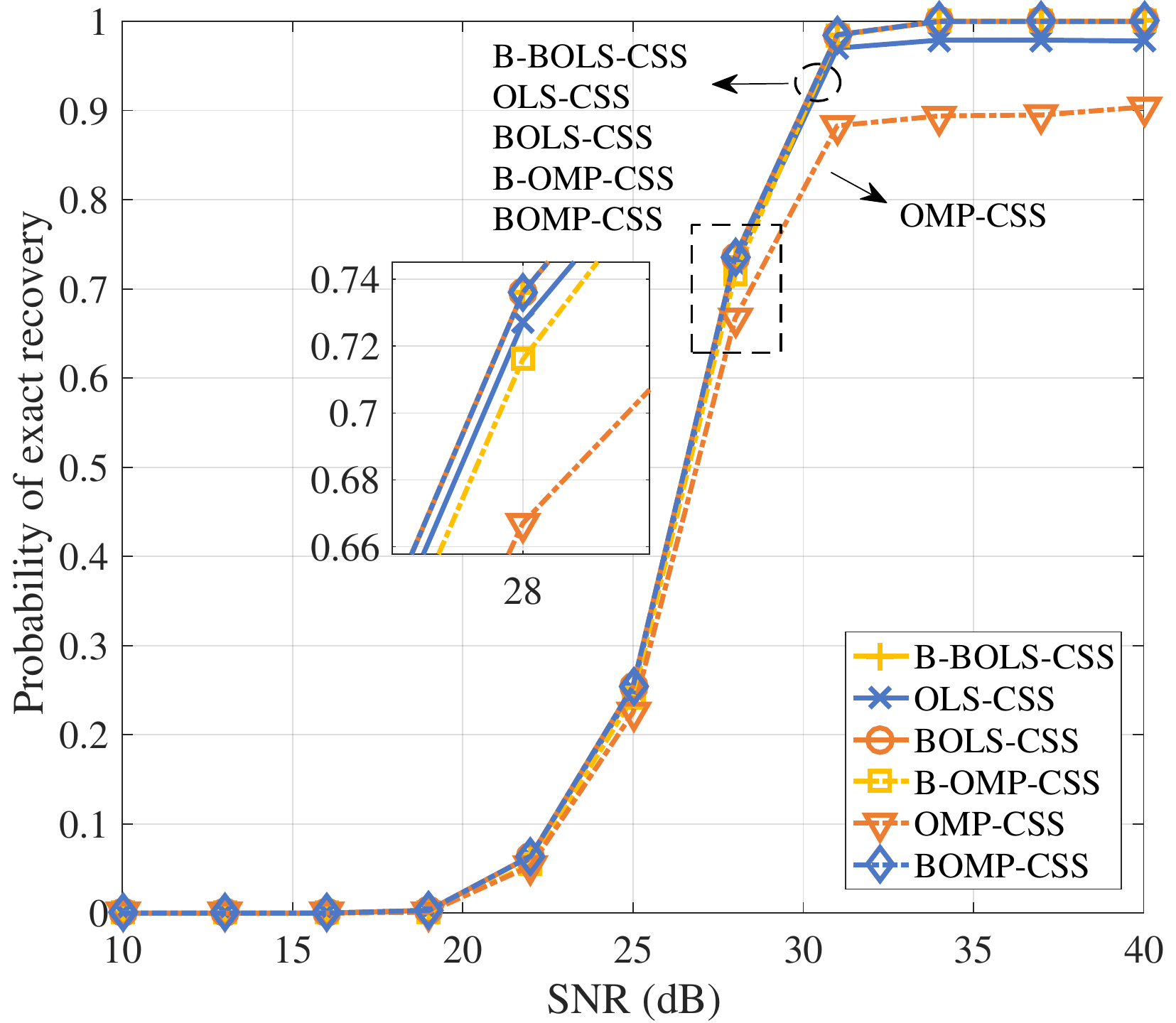}\\
  \caption{Probabilities of exact recovery versus SNR (dB) using hybrid measurement matrix with $m=256$, $n=512$, $d=8$ and the entries of the sparse spectrum satisfy $\mathcal{N}(1,0.01)$.}\label{measurement2snr256512d8model2}
\end{figure}

Figs. \ref{measurement2snr256512d8} and \ref{measurement2snr256512d8model2} give the recovery performance when the number of measurements and block length are doubled to 256 and 8, respectively.
When compared with Figs. \ref{measurement2snr128512d4} and \ref{measurement2snr128512d4model2}, the similar conclusions can be obtained in Figs. \ref{measurement2snr256512d8} and \ref{measurement2snr256512d8model2}.

\section{Conclusion}
In this paper, we propose a B-BOLS-CSS algorithm to address the CSS challenge arose with unknown prior information, e.g., sparsity and noise information. Our theoretical and empirical work demonstrate that the proposed B-BOLS-CSS algorithm performs well in wideband spectrum sensing when these prior information is absent. Our results provide several improvements over previous work on MIP-based recovery condition analyses, and they significantly reduce the required SNR bound for the implementation of the blind recovery algorithm. In other words, B-BOLS-CSS, which exploits the block structure and the fast greedy property, is an effective and robust CSS algorithm.

\section*{Acknowledgment}
This work was supported by the National Natural Science
Foundation of China (61871050), US National Science
Foundation (2003211, 2128596, and 2136202) and Virginia Research Investment Fund (CCI-223996).

\appendix

\section{Proof of Lemma \ref{lemma8}}
\label{proofoflemma8}

We first present the proof for the lower bound of $\lambda_{\min}$. It suffices to prove that the matrix $\mathbf{D}_0^T\mathbf{D}_0-\lambda \mathbf{I}$ is nonsingular under the condition $\lambda<1-(k-1)d\mu_B$ when $(k-1)d\mu_B<1$. The proof is equivalent to proving that for any nonzero vector $\mathbf{r}=(r_1,r_2,\cdots,r_{kd})^T\in \mathcal{R}^{kd}$, $(\mathbf{D}_0^T\mathbf{D}_0-\lambda \mathbf{I})\mathbf{r}\neq0$. Without loss of generality, we assume $||\mathbf{r}[1]||_2\geq||\mathbf{r}[2]||_2\geq\cdots\geq||\mathbf{r}[k]||_2$. The $\ell_2$-norm of the first block of $(\mathbf{D}_0^T\mathbf{D}_0-\lambda \mathbf{I})\mathbf{r}$ satisfies
\begin{equation}\label{thefirstblock}
\begin{aligned}
&||\{(\mathbf{D}_0^T\mathbf{D}_0-\lambda \mathbf{I})\mathbf{r}\}[1]||_2\\
=&||(\mathbf{D}_0[1]^T\mathbf{D}_0[1]-\lambda \mathbf{I})\mathbf{r}[1]+\sum_{i=2}^{k}\mathbf{D}_0[1]^T\mathbf{D}_0[i]\mathbf{r}[i]||_2\\
\geq&||\mathbf{D}_0[1]^T\mathbf{D}_0[1]\mathbf{r}[1]||_2-\lambda|| \mathbf{r}[1]||_2-d\mu_B(\sum_{i=2}^k||\mathbf{r}[i]||_2)\\
\overset{(a)}{\geq}&(1-\lambda)||\mathbf{r}[1]||_2-d\mu_B(\sum_{i=2}^k||\mathbf{r}[i]||_2)\\
>&(k-1)d\mu_B||\mathbf{r}[1]||_2-d\mu_B(\sum_{i=2}^k||\mathbf{r}[i]||_2)\\
\geq&0,
\end{aligned}
\end{equation}
where $(a)$ is obtained because of $\nu=0$.
Afterwards, $(\mathbf{D}_0^T\mathbf{D}_0-\lambda \mathbf{I})\mathbf{r}\neq0$ and we obtain $\lambda_{\min}\geq1-(k-1)d\mu_B$. $\lambda_{\max}\leq1+(k-1)d\mu_B$ is obtained similarly.

\section{Proof of Theorem \ref{theorem8}}
\label{proofoftheorem8}

The proof of the theorem contains three steps: (1) BOLS chooses a correct nonzero entry during each iteration; (2) BOLS does not stop in the $t$-th step $(t<k)$; (3) BOLS stops after $k$ iterations.

The condition of the first step follows from Theorem \ref{theorem6}. After some simple calculations, we have
\begin{equation}\label{directcalculations}
\begin{aligned}
&\frac{||\mathbf{x}_{0\backslash\mathbf{S}}||_2}{\sigma}=\sqrt{\sum_{i\in0\backslash\mathbf{S}}\frac{|\mathbf{x}_i|^2}{\sigma^2}}=\sqrt{\sum_{i\in0\backslash\mathbf{S}}m\times {\rm SNR}_i}\\
>&\frac{2\sqrt{k-t}(2-(k-\mathcal{B})d\mu_B)\sqrt{d}\sqrt{m+2\sqrt{m\log m}}}{(1-(k-1)d\mu_B)(2-(k-\mathcal{B})d\mu_B-2\mathcal{B}kd\mu_B)}.
\end{aligned}
\end{equation}
This means BOLS chooses a correct block if
\begin{equation}\label{snr1}
\begin{aligned}
{\rm SNR}_{\min}
>\bigg(\frac{2(2-(k-\mathcal{B})d\mu_B)\sqrt{d}\sqrt{m+2\sqrt{m\log m}}}{\sqrt{m}(1-(k-1)d\mu_B)(2-(k-\mathcal{B})d\mu_B-2\mathcal{B}kd\mu_B)}\bigg)^2.
\end{aligned}
\end{equation}

Then, for $t<k$, with the probability ${\rm Pr}\{||\mathbf{D}^T\mathbf{P}_t^{\bot}\epsilon||_{2,\infty}\leq\sqrt{d}\xi\mu_B\eta\sigma,||\mathbf{\epsilon}||_2\leq\sqrt{m+2\sqrt{m\log m}}\sigma\}$, we obtain
\begin{equation}\label{discussion1}
\begin{aligned}
\frac{||\mathbf{D}^T\mathbf{r}^t||_{2,\infty}}{||\mathbf{r}^t||_2}
\geq&\frac{\frac{1}{\sqrt{k-t}}||\mathbf{D}_{0\backslash\mathbf{S}}^T\mathbf{P}_{\mathbf{S}^t}^T\mathbf{D}_{0\backslash\mathbf{S}}\mathbf{x}_{0\backslash\mathbf{S}}||_2-\sqrt{d}\xi\mu_B\eta\sigma}{||\mathbf{P}_{\mathbf{S}^t}^{\bot}\mathbf{D}_{0\backslash\mathbf{S}}\mathbf{x}_{0\backslash\mathbf{S}}||_2+\sigma\sqrt{m+2\sqrt{m\log m}}}\\
\geq&\frac{\frac{(1-(k-1)d\mu_B)}{\sqrt{k-t}}||\mathbf{x}_{0\backslash\mathbf{S}}||_2/\sigma-\sqrt{d}\xi\mu_B\eta}{(1+(k-1)d\mu_B)||\mathbf{x}_{0\backslash\mathbf{S}}||_2/\sigma+\sqrt{m+2\sqrt{m\log m}}}\\
\geq&\sqrt{d}\xi\mu_B,
\end{aligned}
\end{equation}
with the ${\rm SNR_{\min}}$ satisfying
\begin{equation}\label{discussion2}
\begin{aligned}
{\rm SNR}_{\min}
>\frac{(\sqrt{d}\xi\mu_B\sqrt{m+2\sqrt{m\log m}})^2}{m(1-(k-1)d\mu_B-\sqrt{kd}\xi\mu_B(1+(k-1)d\mu_B))^2}.
\end{aligned}
\end{equation}

For $t=k$, with the probability ${\rm Pr}\{||\mathbf{D}^T\mathbf{P}_{\mathbf{S}^t}^{\bot}\mathbf{\epsilon}||_{2,\infty}\leq\sqrt{d}\xi\mu_B\eta\sigma,||\mathbf{P}_{\mathbf{S}^t}^{\bot}\mathbf{\epsilon}||_2\geq\eta\sigma\}$, we obtain $\frac{||\mathbf{D}^T\mathbf{r}_k||_{2,\infty}}{||\mathbf{r}_k||_2}\leq\sqrt{d}\xi\mu_B$.

Finally, the probability in (\ref{snrprob}) can be obtained by using $k<\mathcal{C}$.


\bibliographystyle{IEEEtran}
\bibliography{signalprocessing}

\end{document}